\documentclass[final,12pt]{article}
\usepackage[left=1in, top=1in, bottom=1in, right=1in]{geometry}

\usepackage{latexsym,amssymb,amsmath,color}
\usepackage{graphicx}
\usepackage[colorlinks=true]{hyperref}
\hypersetup{urlcolor=blue, citecolor=red}
\usepackage{cite}
\usepackage[table]{xcolor}

\newcommand{\R}{\mathbb{R}}

\newcommand{\be}{\begin{equation}}
\newcommand{\ee}{\end{equation}}
\newcommand{\bea}{\begin{eqnarray}}
\newcommand{\eea}{\end{eqnarray}}

\def\1{{\rm 1}}

\def\s1{^{\rm (1)}}

\renewcommand{\vec}[1]{{\mathchoice
                     {\mbox{\boldmath$\displaystyle{#1}$}}
                     {\mbox{\boldmath$\textstyle{#1}$}}
                     {\mbox{\boldmath$\scriptstyle{#1}$}}
                     {\mbox{\boldmath$\scriptscriptstyle{#1}$}}}}

\newcommand{\normone}[1]{\left\| {#1} \right\|_{\scriptscriptstyle 1}}

\def\1{{\rm 1}}

\def\s1{^{\rm (1)}}

\newcommand{\Np}{{N_\text{p}}}
\newcommand{\Npc}{{N_\text{PC}}}

\def\longrightharpoonup{\relbar\joinrel\rightharpoonup}
\def\longleftharpoondown{\leftharpoondown\joinrel\relbar}
\def\longrightleftharpoons{
  \mathop{
    \vcenter{
      \hbox{
      \ooalign{
        \raise1pt\hbox{$\longrightharpoonup\joinrel$}\crcr
          \lower1pt\hbox{$\longleftharpoondown\joinrel$}
          }
      }
    }
  }
}

\newcommand{\GG}{G}

\usepackage[table]{xcolor}
\usepackage{booktabs} 
\usepackage{relsize}
\usepackage{bm}
\usepackage{mathrsfs}
\usepackage{amsmath,amssymb}

\usepackage[latin1]{inputenc}
\usepackage{tikz}
\usetikzlibrary{positioning,shapes,arrows,arrows.meta}
\usetikzlibrary{shapes.geometric}
\tikzstyle{decision} = [diamond, minimum width=2.5cm, minimum height=1cm, text centered, draw=black]
\tikzstyle{startstop} = [rectangle, draw, 
   text width=4.5em, text centered, rounded corners]
 \tikzstyle{process} = [rectangle, draw, 
   text width=9.0em, text centered]  
\tikzstyle{line} = [draw, -latex']
\tikzstyle{cloud} = [draw, ellipse, node distance=3cm,
   minimum height=2em]
   \tikzstyle{io} = [trapezium, trapezium left angle=70, trapezium right angle=110, minimum width=2cm, text width=9.5em,minimum height=1cm, text centered, draw, inner sep=0pt]

\usepackage{algorithm,algpseudocode,float}
\usepackage{lipsum}
\makeatletter

\newenvironment{breakablealgorithm}
  {
   \begin{center}
     \refstepcounter{algorithm}
     \hrule height.8pt depth0pt \kern2pt
     \renewcommand{\caption}[2][\relax]{
       {\raggedright\textbf{\ALG@name~\thealgorithm} ##2\par}%
       \ifx\relax##1\relax 
         \addcontentsline{loa}{algorithm}{\protect\numberline{\thealgorithm}##2}%
       \else 
         \addcontentsline{loa}{algorithm}{\protect\numberline{\thealgorithm}##1}%
       \fi
       \kern2pt\hrule\kern2pt
     }
  }{
     \kern2pt\hrule\relax
   \end{center}
  }
  \makeatother
\algdef{SE}[DOWHILE]{Do}{doWhile}{\algorithmicdo}[1]{\algorithmicwhile\ #1}%

\begin{document}

\thispagestyle{empty}
\begin{center}
\textsc{
Sensitivity-driven adaptive construction of reduced-space surrogates
}

\bigskip 
\bigskip 

Manav Vohra$^{1}$, Alen Alexanderian$^{2}$, Cosmin Safta$^{3}$, Sankaran Mahadevan$^{1}$

\bigskip
\bigskip

\normalsize
$^1$Department of Civil and Environmental Engineering\\
Vanderbilt University\\
Nashville, TN 37235\\

\bigskip

$^2$Department of Mathematics\\
North Carolina State University\\
Raleigh, NC 27695\\

\bigskip

$^3$Sandia National Laboratories\\
Livermore, CA 94550\\

\end{center}

%
%
%

\baselineskip=22pt


\section*{Abstract}
We develop a systematic approach for surrogate model
construction in reduced input parameter spaces.
%
A sparse set of model evaluations in the original input space is used to  
approximate derivative based global sensitivity measures (DGSMs) 
for individual uncertain inputs of the model.
An iterative screening procedure is developed that exploits DGSM estimates in
order to identify the \emph{unimportant} inputs. The screening procedure forms
an integral part of an overall framework for adaptive construction of a
surrogate in the reduced space. The framework is tested for computational
efficiency through an initial implementation in simple test cases such as the
classic Borehole function, and a semilinear elliptic PDE with a random source
term. 
The framework is then deployed for a realistic application from chemical
kinetics, where we study the ignition delay in an H$_2$/O$_2$ reaction
mechanism with 19 uncertain rate constants.  It is observed that significant
computational gains can be attained by constructing accurate low-dimensional
surrogates using the proposed framework.

\clearpage

\section{Introduction}
\label{sec:intro}

The emerging field of uncertainty quantification (UQ) aims at methodologies for
incorporating, characterizing, quantifying, propagating, and reducing the
uncertainties associated with predictive models and simulations.  For
situations involving complex physical models and computationally intensive
simulations, surrogate modeling often provides orders of magnitude speedups in
statistical studies. This is done by replacing repeated evaluations of
computationally expensive models by inexpensive evaluations of a surrogate
model.  Thus, an efficient approach to construction of surrogate models is of
central importance in enabling efficient uncertainty quantification for
computationally intensive models.

Commonly used surrogate modeling approaches use polynomial chaos expansions
(PCEs)~\cite{Xiu:2002,Ghanem:2003,Olivier:2010}, multivariate adaptive
regression splines (MARS)~\cite{friedman93}, Gaussian processes
(GPs)~\cite{Rasmussen:2004}, or Kriging~\cite{Stein:2012}.  Many real-world
applications involve a large number of model inputs. This makes
the construction of surrogate models difficult or impossible in some cases.
However, in many situations, the variability in model observables of interest
is sensitive to only a small subset of the uncertain inputs.  Hence,
identifying model inputs that are inessential to variability in model output is
a key step that can help reduce the input parameter dimension and hence the effort
associated with surrogate model construction. 



Variance based global sensitivity analysis based on Sobol' indices~\cite{Sobol93,
Sobol:2001,
Owen13,
SaltelliRattoAndresEtAl08} provides
insight into the relative contributions of the uncertain model inputs to the
uncertainty in predictions. Specifically, such analysis can help reduce the
dimensionality of the problem.  Computing Sobol' indices, however, is a
computationally demanding task.  
Availability of a surrogate model typically
enables efficient computation 
of Sobol' indices~\cite{Sudret08,CrestauxLeMaitreMartinez09,BlatmanSudret10,HartAlexanderianGremaud17,Sargsyan17}. This has enabled performing global sensitivity analysis
on a wide range of applications including in ocean 
modeling~\cite{AlexanderianWinokurSrajEtAl12,LiIskandaraniLeHenaffEtAl16},
geosciences~\cite{Namhata2016OladyshkinDilmoreEtAl16,deman2016,SaadAlexanderianPrudhommeEtAl17},
and chemical kinetics~\cite{DegasperiGilmore08,navarro2016global,Vohra:2014}
to name a few.

While surrogate models provide an efficient way of computing sensitivity
indices, constructing them in the case of models with high-dimensional inputs
can be as expensive as computing the Sobol' indices via sampling. 
In this article, we propose a practical and efficient
approach to address this commonly observed ``chicken-and-egg'' problem
in surrogate modelling for engineering applications.
Specifically, we reduce
the dimensionality of the input space using derivative-based global sensitivity
analysis~\cite{Sobol:2009,Sobol:2010,Lamboni:2013,Kucherenko:2009,Kucherenko:2016},
which enables a tractable approach for global sensitivity
analysis~\cite{Kucherenko:2016}. The links between derivative based
global sensitivity measures (DGSMs) and total Sobol'
indices~\cite{Sobol:2009,Kucherenko:2009,Kucherenko:2016} provide a strong
basis for their use in identifying unimportant parameters. In addition
to the construction of an efficient surrogate in the reduced space,
dimension reduction highlights key features of the input-output relationship
encapsulated by the model, and allows for an efficient approach to calibration
of the important inputs.


\paragraph{Our approach}
We present a strategy for identifying and screening uncertain model parameters
that are significantly less important than the rest, thereby reducing the
dimensionality of the problem and enabling the construction of a reduced-space
surrogate (RSS).  Our approach combines DGSMs and surrogate modeling in an
iterative manner.  To make optimum use of computational resources, batches of
model evaluations are performed iteratively, and convergence of our DGSM based
screening metric is tested successively. Moreover, a series of verification steps
incorporated in our method enable monitoring the accuracy of parameter-screening
and the resulting surrogate model.  Our approach is agnostic to the choice of
methodology for constructing the surrogate. However, in the present work, we
rely on sparse polynomial chaos expansions (PCEs) 
to demonstrate the suitability of the proposed strategy.  


\paragraph{Contributions}
The contributions of this article are as follows: (i) We establish a robust and
practical framework for dimension reduction and surrogate modeling using
derivative-based global sensitivity measures. Our
approach is general in that it is applicable to a wide range of applications.
(ii) We present comprehensive numerical results demonstrating the viability of
our strategy using motivating applications: the classic borehole
function, and a semilinear elliptic PDE.  (iii) We 
deploy our strategy in an application problem from chemical kinetics with 19
uncertain parameters. The problem is studied in multiple regimes.
It is shown that the 19-parameter problem can be efficiently reduced to a 3- 
or 4-dimensional problem.

\paragraph{Paper outline}
This article is structured as follows. In section~\ref{sec:bg}, we provide a
brief introduction to DGSMs as well as the polynomial chaos methodology used in
the present work.  In section~\ref{sec:method}, we present our proposed
approach, where we also provide a detailed numerical algorithm and a flow
diagram to aid practitioners in implementing the presented framework.
Section~\ref{sec:examples} is devoted to numerical examples
examining various aspects of our approach. This is followed by implementation
of our framework in a H$_2$/O$_2$ chemical kinetics problem, in section~\ref{sec:app}.
Finally, concluding remarks are provided in section~\ref{sec:disc}.


\bigskip
\bigskip

\section{Background}
\label{sec:bg}

In this section, we introduce the notations used in the rest of
the article, and present the requisite background material on 
derivative-based global sensitivity measures and surrogate modeling 
using polynomial chaos expansions.

\subsection{Derivative-based global sensitivity analysis}  
\label{sub:dgsm}

Let $\GG$ be a mathematical model that is a function of $\Np$ uncertain 
inputs, $\theta_1, \theta_2, \ldots, \theta_\Np$. The goal of sensitivity analysis
is measuring the influence of each component of the input vector 
$\bm{\theta} = \begin{bmatrix}\theta_1 & \theta_2 & 
\ldots & \theta_\Np\end{bmatrix}^T$ on the model output. 
In the present work, we consider the case where the inputs are statistically 
independent. 

Derivative-based global sensitivity analysis is performed by 
computing derivative based global sensitivity measures (DGSMs)~\cite{Sobol:2009} 
for each uncertain parameter in the model. 
Specifically, we consider the following DGSMs, 
\be
\mu_i = 
\mathbb{E}\left[\left(\frac{\partial \GG(\bm{\bm{\theta}})}{\partial \theta_i}\right)^{2}\right], \quad i = 1, \ldots, \Np.
\label{eq:mu}
\ee
Here $\mathbb{E}$ denotes expectation over the uncertain parameters.
Notice that this formulation assumes that the function $\GG$ is differentiable
with respect to $\theta_i$, $i = 1, \ldots, \Np$.

If an analytic expression for $\GG$ is available, the derivative in the above
expression can be computed directly. In real-world applications, 
however, $\GG$ is
often defined in terms of a solution of a mathematical model.  In the
present work, we consider a generic computational model and only assume that
the model output depends differentiably on the parameter $\bm{\theta}$. A
simple approach to computing the gradient is to use finite-differences: 
\be
\frac{\partial \GG(\bm{\theta})}{\partial \theta_i} 
\approx
\frac{\GG(\theta_1,\ldots,\theta_{i-1},
\theta_i+\Delta\theta_i,
\theta_{i+1},\ldots,\theta_d) - 
\GG(\bm{\theta})}{\Delta\theta_i}, \quad i = 1, \ldots, \Np. 
\label{eq:partial}
\ee
Then,~\eqref{eq:mu} can be evaluated by Monte Carlo (MC) sampling in
the uncertain parameter space. 
The total number of model realizations or function evaluations
needed to
compute $\mu_i$ for a function $G$ of $\Np$ random inputs and using $N$ samples is
therefore, $N\times(\Np+1)$. 
It is noted in previous studies~\cite{Kucherenko:2016,Kucherenko:2009},
and also
observed in the numerical experiments in the present work, that a modest MC sample size is often sufficient for computing~\eqref{eq:mu} with
reasonable accuracy to identify the unimportant inputs. Moreover, the
computational efficiency for estimating $\mu_i$ can be enhanced by using
techniques such as automatic differentiation
~\cite{Kiparissides:2009} or adjoint-based gradient computation~\cite{jameson1988aerodynamic,gunzburger2003perspectives,Borzi2011,AlexanderianPetraStadlerEtAl17}. 

Consider the total 
Sobol' sensitivity index~\cite{Sobol:2001},
\be
\mathcal{T}(\theta_i) = 1 - 
\frac{\mathbb{V}[\mathbb{E}(\GG|\bm{\theta}_{\sim i})]}{\mathbb{V}(\GG)},
\label{eq:total}
\ee
where $\bm{\theta}_{\sim i}$ is the random vector $\bm\theta$ with $i$th component removed, 
and $\mathbb{V}$ denotes the variance. The total Sobol' index quantifies the total contribution 
of $\theta_i$ to variance of the model $\GG$. Components of $\bm\theta$ with small 
total Sobol' index can be considered inessential and can be fixed at nominal values. However, 
computing the total Sobol' index is a computationally expensive task for expensive-to-evaluate 
models with large number of uncertain parameters. Fortunately, 
for parameters with continuous distributions, an upper bound on $\mathcal{T}_i$  
can be expressed in terms of $\mu_i$ as follows: 
\be
\mathcal{T}(\theta_i) \leq \frac{\mathcal{C}_i\mu_i}{\mathbb{V}(\GG)},
\label{eq:bound}
\ee
where $\mathcal{C}_i$ is the corresponding ``Poincar\'e constant''
and
$\mathbb{V}(\GG)$ is the total variance of the model
output~\cite{Lamboni:2013}.  The upper bound in the above inequality is
proportional to the product of $\mathcal{C}_i$ and $\mu_i$.  For the purpose of
parameter screening as discussed later in Section~\ref{sec:method}, we consider
a normalized product, $\widehat{\mathcal{C}_i\mu_i}$ to ensure that it lies
between 0 and 1:
\be
\widehat{\mathcal{C}_i\mu_i} = \frac{\mathcal{C}_i\mu_i}{\sum_i \mathcal{C}_i\mu_i}.
\label{eq:cmu}
\ee
The Poincar\'e constant, $\mathcal{C}_i$ is specific to the probability
distribution of $\theta_i$. For $\theta_i \sim \mathcal{U}[a, b]$,
$\mathcal{C}_i = (b-a)^{2}/\pi^2$, and for  
$\theta_i \sim \mathcal{N}(\mu,\sigma^2)$, we have $\mathcal{C}_i = \sigma^2$. 
Here $\mathcal{N}(\mu,\sigma^2)$ denotes a normal distribution with mean $\mu$
and variance $\sigma^2$, and $\mathcal{U}[a,b]$ denotes a uniform distribution
on the interval $[a, b]$.

%

\subsection{Polynomial chaos expansion}

We consider models with $\Np$ random inputs, 
$\theta_1, \ldots, \theta_\Np$ that are modeled
as statistically independent random variables. The 
variables $\theta_i$ will take in physically meaningful
ranges; it is common to parameterize input uncertainties
with canonical random variables $\xi_1, \ldots, \xi_\Np$,
which can be then shifted and scaled to obtain the corresponding $\theta_i's$.
Typical choices for distribution of $\xi_i$ include standard normal 
and uniform distribution on the interval $[-1, 1]$.
Let 
\[
   \bm{f}(\bm{x}) = \prod_{i=1}^\Np f_i(x_i), \quad \bm{x} \in \mathbb{R}^\Np
\]
where $f_i$ are probability density functions of $\xi_i$, $i = 1, \ldots, \Np$.

Consider a square integrable random variable $\GG:\R^\Np \to \R$; 
i.e.,
$\int_{\mathcal{D}} \GG(\bm{\xi})^2 \, \bm{f}(\bm{\xi})d\bm{\xi} < \infty$,
where $\mathcal{D}$ is the support of the distribution law of the random vector
$\bm{\xi}$. 
%
The PCE of
$\GG$ is a mean-square 
convergent series expansion~\cite{Xiu:2002,Ghanem:2003,Olivier:2010} of the form:
\be
\GG(\bm\xi) = \sum_{k=0}^\infty c_k\Psi_k(\bm{\xi}),
\ee
where $\Psi_k$'s form a multivariate orthogonal polynomial
basis---orthogonal with respect to the joint probability distribution of $\bm{\xi}$.
In practice, a truncated expansion is used.  Moreover, in applications, $\GG$
is a mathematical model of interest that takes a parameter vector $\bm{\theta}$
(with components in physically meaningful ranges) as input. Therefore, we 
write the truncated PC representation of a model $\GG$ as follows:
\be 
\GG(\bm\theta) \approx \GG^{\mbox{\tiny PC}}(\bm\theta) := \sum_{k=0}^{\Npc}
c_k\Psi_k (\bm\xi(\bm\theta)), 
\ee
where $\bm\xi(\bm\theta)$ is found by a simple linear transformation.

Computational strategies available for estimating the PC coefficients
($c_k$'s) typically involve techniques based on projection or regression.
Projection-based methods consider the orthogonal projection of 
$\GG$ on the PC basis $\{\Psi_k\}_{k=0}^\Npc$ and compute
the resulting expansion coefficients via quadrature~\cite{Olivier:2010}.
Regression-based methods such as least angle regression (LAR)~\cite{Efron:2004}, and least absolute shrinkage
and selection operator (LASSO)~\cite{Tibshirani:1996} aim to construct a sparse PCE~\cite{Blatman:2008}
by solving a penalized least-squares problem. Specifically in the case of LAR, a penalty term comprising the $\ell_1$-norm of the PC coefficients is used:
\be
\hat{\bm{c}} = \mbox{argmin}~\mathbb{E}_{\bm\theta}
\left[\left(\sum_{k=0}^\Npc c_k \Psi_k(\bm\xi(\bm\theta)) -
\GG(\bm{\theta})\right)^{2}\right]  + \lambda\normone{\bm{c}},
\label{eq:reg}
\ee
where $\normone{\bm{c}}$ = $\sum_{k=0}^\Npc |c_k |$.
The penalty term forces the minimization towards sparse coefficient vectors
resulting in sparse PC representations.  In this work, we construct sparse PCEs
with LAR using UQLab~\cite{Marelli:2014}, a general purpose uncertainty
quantification software developed at ETH Zurich.

\bigskip
\bigskip

\section{Methodology}
\label{sec:method}

%

In this section, we outline the underlying framework for \emph{adaptively}
constructing a reduced-space surrogate (RSS) using sensitivity analysis.  The
proposed methodology is described as adaptive since the RSS is constructed only
in situations where it is expected to yield computational dividend as discussed
further below.  The term reduced-space implies that the surrogate is
constructed in a reduced parameter space that sufficiently captures the uncertainty in the
model output. We begin by outlining an algorithm for parameter screening
to assess the importance of individual parameters for potential dimension
reduction and construction of an RSS. The overall adaptive framework 
that incorporates parameter screening as an integral step is thereafter
presented. Finally, we present metrics used for assessing the
convergence and accuracy of the RSS followed by a brief discussion on
salient features of the proposed framework. 

\textbf{Parameter screening.}
In the proposed framework, we
adopt a novel approach for constructing an RSS based on estimating the
upper-bound $\widehat{\mathcal{C}_i\mu_i}$, given in~\eqref{eq:bound}, on 
total Sobol' index ($\mathcal{T}(\theta_i)$) for each 
parameter $\theta_i$; the \emph{screening metrics}, 
$\{\widehat{\mathcal{C}_i\mu_i}\}_{i=1}^\Np$,  
are used to identify
parameters that are relatively unimportant. 


An initial set of $n_1$ samples is used to obtain a rough estimate
of the metric.  
Based on the associated metric value, an initial
rank ($\mathcal{R}_i^{old}$) is assigned to each parameter. At each iteration,
a new set of samples is generated based on the joint probability distribution
of $\bm{\theta}$ and corresponding model
output at each sample point is computed. The new set of gradient evaluations
combined with prior evaluations is used to update parameter ranks. Additionally,
deviation in the derivative-based sensitivity measure between successive iterations
normalized by the measure in the previous iteration is recorded for each parameter.
The iterative process is continued until parameter ranks between successive iterations
are observed to be consistent as well as the maximum deviation among all parameters
($\Delta\mu_s$) is below a certain tolerance ($\tau$). The amount of computational
effort associated with the screening process is limited
by the choice of maximum number of iterations, $s_\text{max}$. 

Key inputs to the screening procedure
are as follows: (1) 
a limiting value $\tau$ of the maximum relative change in the sensitivity
measure between successive iterations; (2) 
a limiting ratio $\tau_\text{screen}$ of the sensitivity metric 
relative to its maximum value;
(3) a real number $\beta \in (0, 1)$ to 
guide the number of new samples $\lceil \beta n_1 \rceil$ 
at each iteration ($\lceil \beta n_1
\rceil$ 
is the smallest integer greater than or equal to $\beta n_1$); 
(4) a set of samples $\{ \bm{\theta}_k \}_{k = 1}^{n_1}$ 
for the initial screening step in
the algorithm and the corresponding gradient evaluations 
$\{ \bm{g}^k \}_{k=1}^{n_0}$, where
$\bm{g}^k = \nabla_{\bm{\theta}} G(\bm{\theta}_k)$. 
The outputs are the set of active
indices $\mathcal{I}_\text{active}$ corresponding to the 
\emph{important parameters}, the total number of available model 
evaluations $N_\text{total}$, and the enriched set of gradient evaluations 
$\{ \bm{g}^k\}_{k=1}^{N_\text{total}}$. 
A general methodology for parameter screening is provided below in
Algorithm~\ref{alg:screen}.

\bigskip
\begin{breakablealgorithm}
\renewcommand{\algorithmicrequire}{\textbf{Input:}}
\renewcommand{\algorithmicensure}{\textbf{Output:}}
  \caption{Parameter screening with DGSMs: A generalized approach.}
  \begin{algorithmic}[1]
\Require $\tau > 0$, $\tau_\text{screen} > 0$, $s_\text{min} \geq 1$,
$s_\text{max} \geq 1$, $\beta > 0$, $\{ \bm{\theta}_k \}_{k = 1}^{n_1}$, $\{ \bm{g}^k \}_{k=1}^{N_\text{total}}$. 
\Ensure $\mathcal{I}_\text{active}$, $\{ \bm{g}^k \}_{k=1}^{N_\text{total}}$, $N_\text{total}$. 
    \Procedure{Screening}{}
      \State Compute $\bm{g}^k = \nabla_{\bm{\theta}}G(\bm\theta_k)$, 
             $k~=~N_\text{total}+1, \ldots, N_\text{total}+n_1$. 
      \State $N_\text{total} = N_\text{total} + n_1$
      \State Compute 
      $\mu_{1, i} = \frac{1}{N_\text{total}} \sum_{k = 1}^{N_\text{total}} (g^k_i)^2$
      \State Compute $\nu_i = \widehat{\mathcal{C}_i\mu_{1,i}}$, for each $\theta_i$, 
             $i = 1, \ldots, \Np$. 
      \State Determine initial ranks: 
            let $\mathcal{R}^{old} = \{ \nu_{i_1}, \nu_{i_2}, \ldots, \nu_{i_\Np}\}$ such that 
\[
   \nu_{i_1} \geq \nu_{i_2} \geq \cdots \geq \nu_{i_\Np}. 
\]
      \State Set $s$ = 1 and $\mathrm{done} = \mathrm{false}$.
      \While {$\mathrm{done} == \mathrm{false}$ \textbf{AND} $s \leq s_\text{max}$} 
        \State $s = s + 1$.
        \State Draw $n_s = \lceil \beta n_1 \rceil$ new samples 
                  $\bm{\theta}_k$, $k = n_{s-1} + 1, \ldots, n_{s-1} + n_s$
       \State $N_\text{total} = N_\text{total} + n_s$.
        \State Compute $\bm{g}^k = \nabla_{\bm{\theta}}G(\bm\theta_k)$,
             $k = n_{s-1}+1, \ldots, n_{s-1}+n_s$.

        \State Compute $\{ \mu_{s,i} \}_{i=1}^\Np$ using the augmented sample 
               $\{\bm{g}_k \}_{k = 1}^{N_\text{total}}$.
        \State Compute $\nu_i = \widehat{\mathcal{C}_i\mu_{s,i}}$, $i = 1, \ldots \Np$.
        \State Determine new ranks $\mathcal{R}^{new}$ based on $\{\nu_i\}_{i=1}^\Np$. 
        \State Compute $\displaystyle\Delta\mu_s = \max_{1\leq i \leq \Np}
               \left(\frac{|\mu_{s,i} - \mu_{s-1,i}|}{ \mu_{s-1,i}}\right)$.
      \If {$\mathcal{R}^{\tiny{new}} = \mathcal{R}^{\tiny{old}}$ {\bf AND}  $\Delta\mu_s \leq \tau$
                {\bf AND} $s \geq s_\text{min}$}
         \State $\mathrm{done} = \mathrm{true}$
      \Else
          \State Set $\mathcal{R}^{old} = \mathcal{R}^{new}$
      \EndIf
    \EndWhile
    \State $\mathcal{I}_\text{active} = \{ i \in \{1, \ldots, \Np\} : \displaystyle\frac{\nu_i}
        {\|\bm{\nu}\|_\infty} > \tau_\text{screen}\}.$
    
    \EndProcedure
  \end{algorithmic}
  \label{alg:screen}
\end{breakablealgorithm}
\bigskip

\textbf{Adaptive surrogate model construction.}
%
We begin by allocating computational resources for constructing a cross-validation
test suite to be used for assessing the accuracy of the resulting surrogate.
Naturally, the resources allocated for this purpose depend upon the application
as well as total amount of available resources. 
The set of required inputs for parameter screening are initialized,
and model evaluations at $n_1$ random samples in the full-space are computed.
These evaluations are used to construct a surrogate in the full-space (FSS)
using regression-based techniques. If the surrogate is found to be sufficiently
accurate for the given application, the process is terminated. However, it is
likely that a full-space surrogate constructed using a 
small number of model evaluations would not provide a 
faithful representation of the input-output relationship.  

\begin{figure}[htbp]
\tikzset{
    arro/.style={{Square[]->}}
}  

\begin{tikzpicture}[node distance=1.5cm]

\node (start) [startstop] {Start};

\node (val) [process, below of=start, text width=14.5em] {Create a cross-validation
test suite using pre-allocated resources.};

\draw [arro] (start) -- (val);

\node (qoi) [io, below of=val,align=left,yshift=-0.2cm] {Select an appropriate model output};

\draw [arro] (val) -- (qoi);

\node (tol) [io, below of=qoi, text width=9em,align=left,yshift=-0.2cm] {Initialize: $\tau$, $\tau_\text{screen}$, 
$s_\text{max}$, $\beta$, $N_\text{total}$ = 0};

\draw [arro] (qoi) -- (tol);

\node (sam0) [process, below of=tol, text width=12.5em, yshift=-0.3cm] {Draw $n_1$ samples $\{ \bm{\theta}_k \}_{k = 1}^{n_1}$ 
     according to $\bm{f(\theta)}$};
     
\draw [arro] (tol) -- (sam0);     



\node (fss) [process, below of=sam0, text width=14.5em, yshift=-0.5cm] {Construct regression-based 
surrogate in full-space (FSS) using ($N_\text{total}$+$n_1$) model evaluations};

\draw [arro] (sam0) -- (fss);   

\node (chk) [process, below of=fss, text width=12.5em, yshift=-0.5cm] {Assess accuracy of FSS using the
validation test suite};

\draw [arro] (fss) -- (chk);   

\node (res) [draw, diamond, aspect=1.5, text width=7.5em, text centered, below of=chk,yshift=-1.5cm] {Is \\ FSS, 
sufficiently accurate$?$};

\draw [arro] (chk) -- (res);  

\node (screen) [process, right of=res, text width=9.5em, xshift=4.5cm] {Parameter Screening};

\draw [arro] (res) -- node[anchor=north] {N} (screen);

\node (dr) [draw, diamond, aspect=1.8, text width=7.5em, text centered, right of=screen,xshift=4cm] {Is $\alpha$
small enough$?$};

\draw [arro] (screen) -- (dr); 

\node (dtol) [io, above of=dr, text width=10.5em,align=left,yshift=3.2cm,inner sep=0.2pt] {\vspace{1mm}Update: $\tau$, 
$\tau_\text{screen}$, $s_\text{max}$, $\beta$; Input: $\{ \bm{g}^k \}_{k=1}^{N_\text{total}}$\vspace{1mm}};

\draw [arro] (dr) -- node[anchor=east,yshift=-0.8cm] {N} (dtol);

\draw [arro] (dtol) |- (sam0); 

\node (ros) [process, below of=dr, text width=11.5em, yshift=-1.5cm] {Construct a reduced-space surrogate (RSS)};

\draw [arro] (dr) -- node[anchor=east,yshift=0.1cm] {Y} (ros);

\node (ver) [process, left of=ros, text width=13.0em, xshift=-5cm] {Test RSS accuracy using
 evaluations at $N_\text{total}$ $\&$ the validation test suite};

\draw [arro] (ros) -- (ver); 

\node (stop) [startstop, below of=chk,yshift=-4.5cm] {Stop};

\draw [arro] (res) -- node[anchor=east,yshift=0.0cm] {Y} (stop);

\draw [arro] (ver) -- (stop); 

\end{tikzpicture}

\caption{Flow-diagram outlining the adaptive strategy for constructing reduced-space
 surrogates.}
\label{fig:flow}
\end{figure}
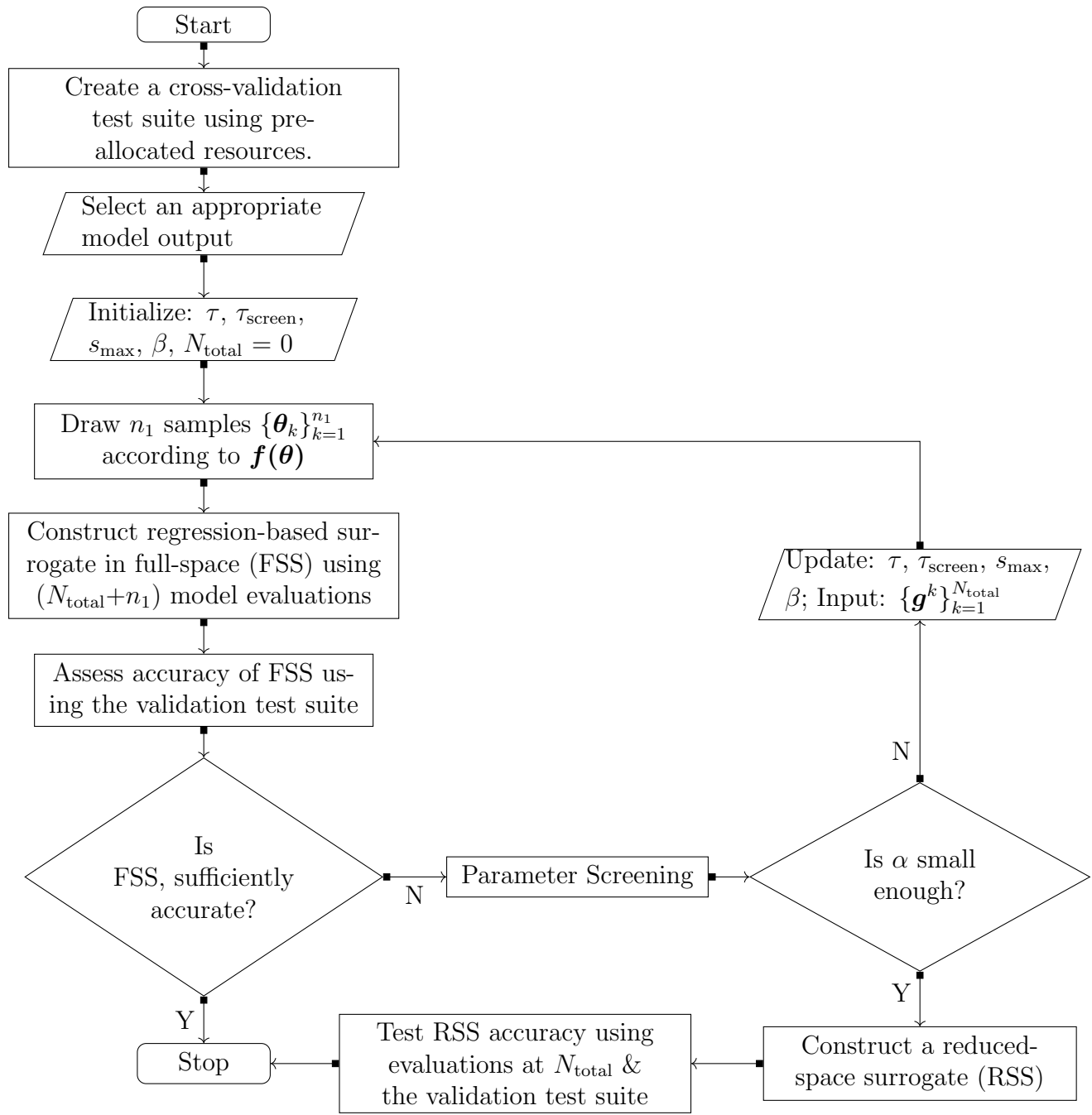

The available set of model evaluations are utilized and further enhanced during
parameter screening as discussed earlier.  At the end of screening, the set of
active indices, $\mathcal{I}_\text{active}$, is used to evaluate $\alpha$, referred
to as the degree of dimension reduction:
\be
\alpha = \frac{|\mathcal{I}_\text{active}|}{N_p},
\label{eq:alpha}
\ee
where $|\mathcal{I}_\text{active}|$ denotes the cardinality of
$\mathcal{I}_\text{active}$.  Scope for dimension-reduction increases as
$\alpha$ decreases.  Hence, if $\alpha$ is considered to be small and
computational gains are expected owing to dimension reduction, the RSS is
constructed and verified for accuracy using a combination of model evaluations
used for screening and those associated with the cross-validation test suite.  On the
other hand, if $\alpha$ is close to 1, the set of inputs required for screening are
 updated as needed, and a
new set of $n_1$ samples and corresponding model evaluations are generated. The
FSS is reconstructed using the enriched set of evaluations and the
aforementioned analysis is repeated as illustrated in the flow-diagram
in Figure~\ref{fig:flow} that shows the overall 
parameter screening and surrogate model construction method.

\textbf{Assessment of the surrogate.}
To assess accuracy of the resulting surrogate, one could estimate the
leave-one-out cross validation error as follows:
\be
\epsilon_{\mbox{\tiny LOO}} = 
\frac{\sum\limits_{i=1}^{N_l}\left(G(\bm{\theta}_i) - G^{\mbox{\tiny {PC}\textbackslash i}}(\bm{\xi(\theta}_i))\right)^2}
{\sum\limits_{i=1}^{N_l}\left(G(\bm{\theta}_i) - \widetilde{\mu}\right)^2},
\label{eq:loo}
\ee
where $N_l$ is the number of training points,
$\widetilde{\mu}~=~\frac{1}{N_l}\sum\limits_{i=1}^{N_l} G(\bm{\theta}_i)$ is
the sample mean of the model response, and $ G^{\mbox{\tiny {PC}\textbackslash
i}}$ is the PC surrogate constructed using all but the $i^{\mbox{\tiny{th}}}$
model realization.  From~\eqref{eq:loo}, it appears that $N_l$ PCEs are needed
to evaluate $\epsilon_{\mbox{\tiny LOO}}$.  However, in practice a modified
formulation for $\epsilon_{\mbox{\tiny LOO}}$~\cite{Blatman:2009}, independent
of $G^{\mbox{\tiny {PC}\textbackslash i}}$ is used; for an easy reference,
see~\cite[Eq.~(1.27)]{Marelli:2014}.  Accuracy of the surrogate could also 
be assessed by evaluating the relative $\ell_2$-norm of the difference in
predictions between the model and the surrogate ($\epsilon_{\mbox{\tiny{L-2}}}$), as follows:
\be
\epsilon_{\mbox{\tiny{L-2}}} = \frac{\left[\sum\limits_{i=1}^{N_v}\left(G(\bm{\theta}_i) - 
G^{\mbox{\tiny {PC}}}(\bm{\xi(\theta}_i))\right)^2\right]^{\frac{1}{2}}}
{\left[\sum\limits_{i=1}^{N_v}\left(G(\bm{\theta}_i)\right)^2\right]^{\frac{1}{2}}}.
\label{eq:l2}
\ee
Here $N_v$ is the number of sampling points in the full parameter space at
which model evaluations are available; this, in the case of an RSS, is 
given by the
augmented set of model evaluations used for validation and screening.  
Accuracy of the surrogate could be further investigated by comparing
probability density functions (PDFs) of the model output based on model
evaluations in the full parameter space and the RSS predictions corresponding
to a large number of samples (say, 10$^6$ for a high-dimensional input space).
However, in realistic problems
involving complex, compute-intensive simulations, constructing the PDF based on model
evaluations would be infeasible.  A practical alternative would be to compare
a (normalized) histogram based on sparse model evaluations with the surrogate-based PDF in
order to gain some insight into the statistical quality of the surrogate. 

\textbf{Discussion on the proposed methodology.}
The amount of computational effort associated with the presented methodology
can be mainly attributed to two steps: I.~Parameter Screening, and 
II.~Constructing a converged RSS. Computational gains are realized in situations
where constructing the surrogate in the full parameter space is more expensive
than the combined cost associated with these steps. Determining the optimal
allocation of computational resources for these steps, however, is not possible
a priori. Hence, in the proposed framework, we exploit the set of model
evaluations used in parameter screening to simultaneously construct the FSS
while keeping a track of its accuracy using the cross-validation test suite.
This would help address situations where significant dimension reduction 
is not possible, and hence, constructing the RSS might result in a 
computational
disadvantage. We suggest using a small number of samples in the initial
screening step (say, $n_1$ = 5) and a relatively large $\tau$ (say,
$\mathcal{O}(10^{-1})$) as a starting point with possible reduction in $\tau$
during subsequent screenings. Pseudo-random sampling approaches such as
Latin hypercube sampling (LHS) and quasi Monte Carlo (QMC) could be used to
generate samples in the input space.

Careful assessment and decision-making is required on whether or not to
proceed with the construction of the RSS at the end of each screening step.
The user should account for factors such as the possible degree of 
dimension reduction, accuracy of the concurrent FSS, and availability of
computational resources. 

The applicability of the proposed framework depends upon the choice of the
model output.  Since the screening metric involves computation of partial
derivatives in the full parameter space, the output must exhibit differentiable
dependence on each parameter. It is therefore likely that for a given
application involving multiple outputs, the RSS can only be constructed for a
selected few, using the approach presented above.  Hence, it is important to
assess the nature of the input-output relationship for a given model prior to
implementing the present framework. 

Additionally, in some cases, the partial
derivative of the output with respect to each uncertain input is not available
analytically. In these cases, one could use finite difference (FD) to approximate
the gradient as illustrated in~\ref{eq:partial}. However, since FD requires
model evaluations at neighboring points, the underlying computational cost is
expected to increase by a factor, $N_p+1$, with $N_p$ being the number of inputs. 
A possible, more efficient alternative to FD, which might be suitable in some cases, involves the use of adjoints for
gradient computation~\cite{Griewank:2008}. In the adjoint approach, each
gradient evaluation requires 
a solution of the state equation (forward solve) and that of the 
corresponding adjoint equation; 
see e.g.,~\cite{jameson1988aerodynamic,gunzburger2003perspectives,Borzi2011}.
The adjoint method, however, requires the availability of an adjoint solver.
Another alternative for efficient gradient computation is the use of 
automatic differentiation~\cite{Kiparissides:2009}.

Using the framework proposed in this section, we aim to construct a reliable
surrogate in the most efficient manner within the constraints of the computational
budget. However, it might be possible that for a given application, the RSS is not
found to be sufficiently accurate. In such a scenario, we suggest enriching the
set of important inputs by incorporating the least unimportant model input
as determined after a series of screening steps, and re-constructing the RSS. 
This process could be repeated depending upon the availability of resources
and the desired accuracy of the surrogate.

\bigskip
\bigskip

\section{Motivating Examples}
\label{sec:examples}

In section~\ref{sec:method}, we presented a framework for constructing
an RSS (if deemed advantageous) by identifying unimportant parameters based on 
estimates of the screening metric, $\widehat{\mathcal{C}_i\mu_i}$,
for individual parameters.
In this section, we motivate the proposed methodology by applying it to two
test problems,
namely, the borehole function, and a semilinear elliptic PDE.
Model evaluations in these test
problems are inexpensive. Therefore, we are able to compare the 
relative importance of model parameters based
on the screening metric (computed by sampling the model) with 
those obtained from converged estimates of $\mathcal{T}(\theta_i)$ (computed
using the surrogate constructed in the
full parameter space (FSS)). Additionally, to illustrate the computational gains, 
we compare
convergence trends as a function of training runs for the RSS and the FSS using 
$\epsilon_{\mbox{\tiny LOO}}$ in Eq.~\ref{eq:loo}. 
Furthermore, as discussed earlier in section~\ref{sec:method}, we compare
PDFs of the model output, obtained using the RSS, the FSS,
as well as true model evaluations, for
the purpose of verification. 

\subsection{Borehole function}

The borehole function~\cite{Morris:1993} is a benchmark reference problem in sensitivity analysis.
It models the discharge of water ($\mathcal{Q}$) through a borehole in terms of
geometrical and physical inputs:
\be
\mathcal{Q} = \frac{\displaystyle
2\pi T_u(H_u - H_l)}{\displaystyle
\ln({r}/{r_w})\Big[1 +
\frac{2LT_u}{
\ln({r}/{r_w})r_w^2K_w} + \frac{T_u}{T_l}\Big]}.
\label{eq:bore}
\ee
The radius of influence, $r$ is fixed at 3698.30 m whereas all other parameters
in the right hand side of~\eqref{eq:bore} are considered 
as uncertain. Hence, $\mathcal{Q} = \mathcal{Q}(\vec{\theta})$ with 
\[
\vec{\theta} = \begin{bmatrix}r_w & L & T_u & H_u & T_l & H_l & 
K_w\end{bmatrix}^T, 
\]
being the vector of
uncertain parameters. Table~\ref{tab:bore} provides distributions of the
uncertain input parameters.

\begin{table}[ht]
\renewcommand*{\arraystretch}{1.2}
\caption{Description and distributions of uncertain inputs in the borehole function
given by~\eqref{eq:bore}.}
\label{tab:bore}
\end{table}
\begin{center}
\begin{tabular}{ll}
\toprule
\textbf{Parameter} & \textbf{Distribution} \\ 
\bottomrule
Borehole radius, $r_w$ (m) & $\mathcal{N}$(0.1,0.016) \\
Borehole length, $L$ (m) & $\mathcal{U}$[1120,1680] \\
Transmissivity of upper aquifer, $T_u$ (m$^2$/yr) & $\mathcal{U}$[63070,115600] \\
Potentiometric head of upper aquifer, $H_u$ (m) & $\mathcal{U}$[990,1110] \\
Transmissivity of lower aquifer, $T_l$ (m$^2$/yr) & $\mathcal{U}$[63.1,116] \\
Potentiometric head of lower aquifer, $H_l$ (m) & $\mathcal{U}$[700,820] \\
Borehole hydraulic conductivity, $K_w$ (m/yr) & $\mathcal{U}$[9855,12045] \\
\bottomrule
\end{tabular}
\end{center}

\begin{figure}[htbp]
 \begin{center}
  \includegraphics[width=0.48\textwidth]{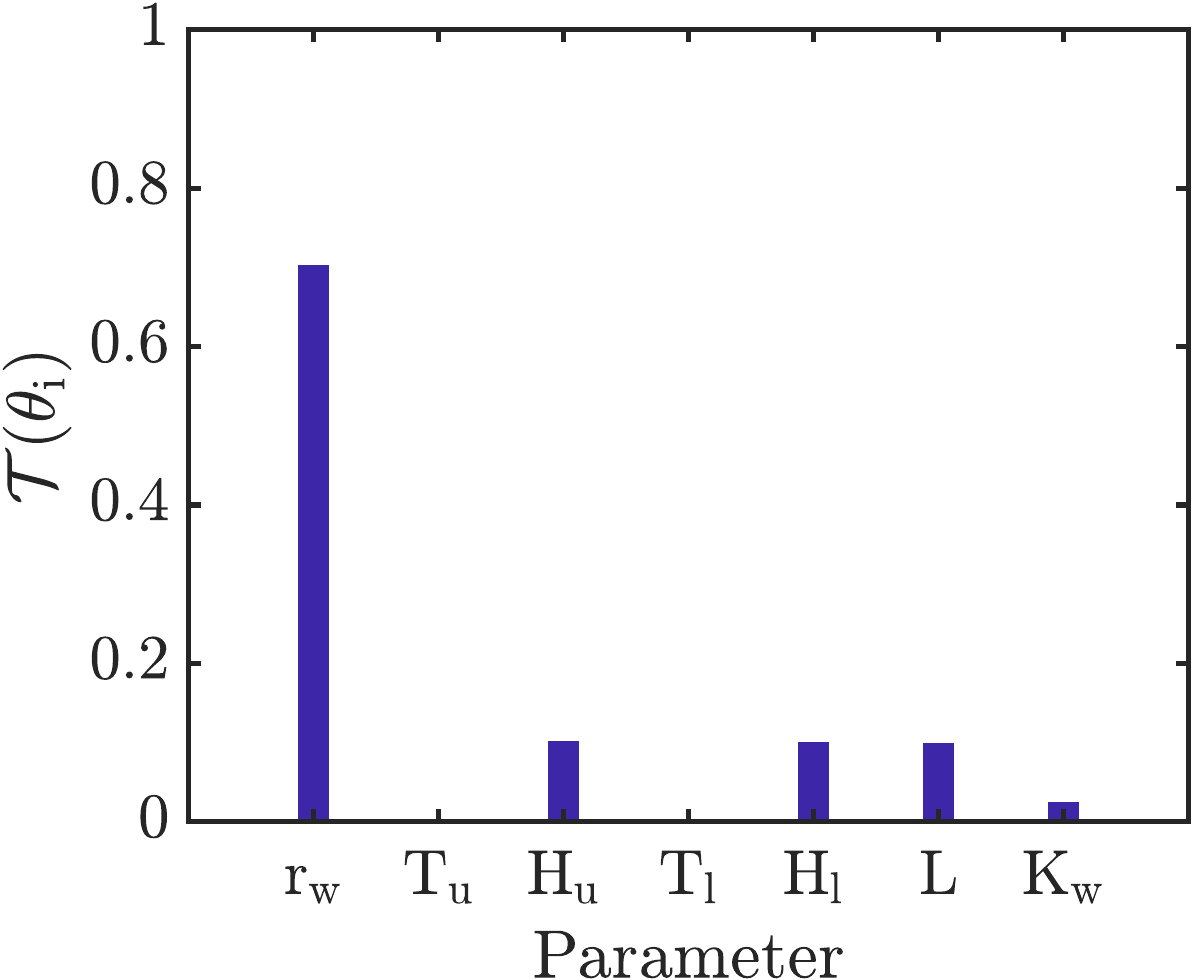}
  \includegraphics[width=0.51\textwidth]{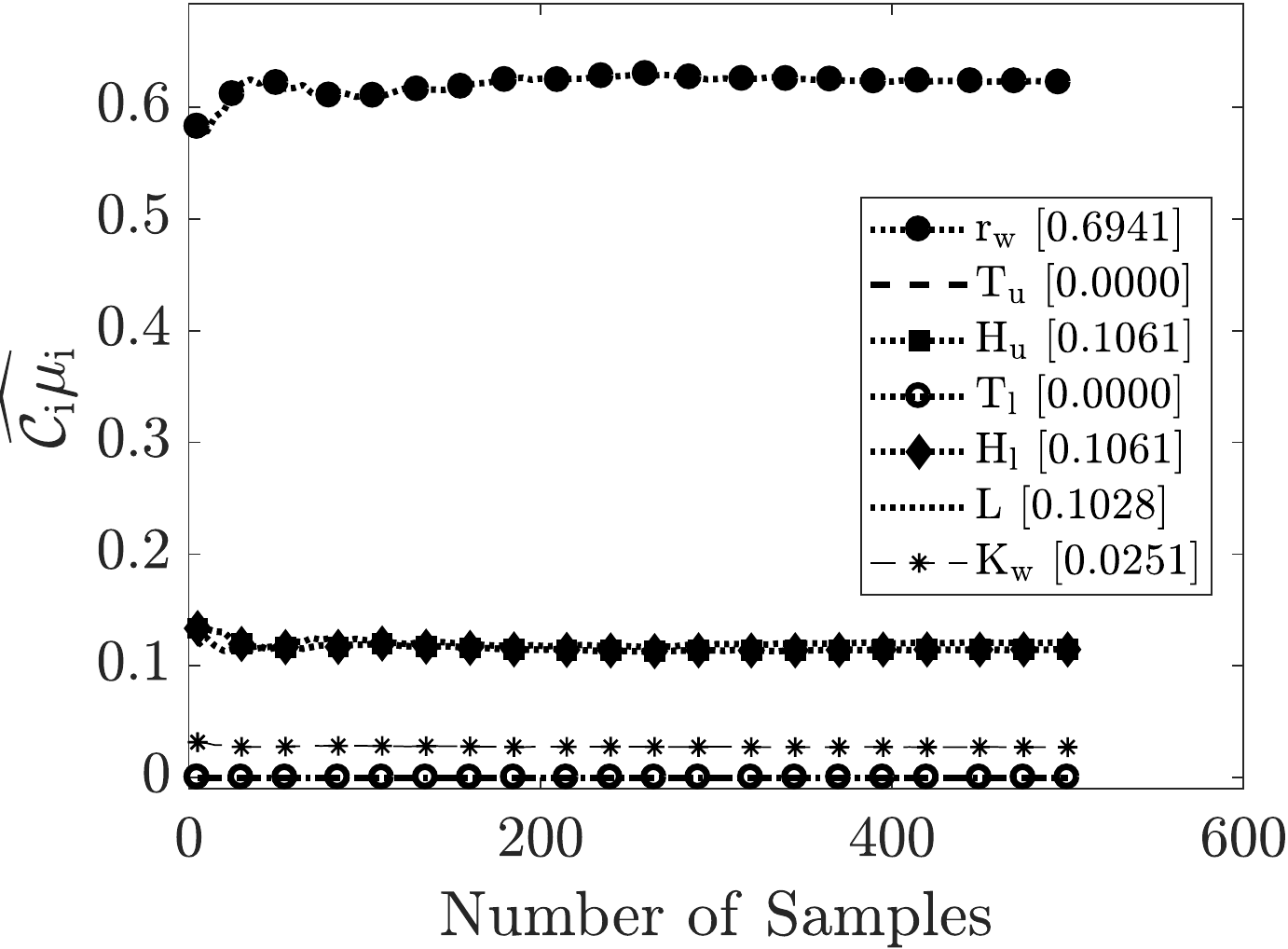}
\caption{
Left: Sobol' total sensitivity index, $\mathcal{T}(\theta_i)$ for uncertain
parameters in the
borehole discharge function in~\eqref{eq:bore}. Right: 
Estimates of the screening metric ($\widehat{\mathcal{C}_i\mu_i}$), plotted
against number of samples. Also included in the legend are estimates of
$\mathcal{T}(\theta_i)$ in each case in the legend.}
\label{fig:sense_bore}
\end{center}
\end{figure}

Cheap function evaluations of the discharge $\mathcal{Q}(\vec{\theta})$ 
enables construction of the FSS with minimal effort. FSS predictions at
a large set of MC samples in the input space are used to obtain converged
estimates of $\mathcal{T}(\theta_i)$. Shown in
Figure~\ref{fig:sense_bore}~(left) are estimates of these indices corresponding to the
uncertain parameters in the borehole function using 10$^6$ MC
samples in the input parameter space. 
These estimates are used to verify fidelity of
parameter screening based on the methodology presented in Section~\ref{sec:method}. 

In Figure~\ref{fig:sense_bore}~(right), we plot estimates of the screening parameter 
$\widehat{\mathcal{C}_i\mu_i}$ for a wide range of the number of 
samples used for approximating $\mu_i$ using~\eqref{eq:mu}.
Estimates for $\widehat{\mathcal{C}_i\mu_i}$ are found to be in excellent agreement
with $\mathcal{T}(\theta_i)$ even when small number of samples (5--10) are used. 
Consequently, the relative importance of uncertain 
parameters in the borehole function is found to be consistent 
with predictions based on the Sobol' index. 
In the considered intervals for the uncertain parameters, it is clear 
that the discharge is insensitive to $T_u$ and $T_l$. 
Moreover, the sensitivity to $K_w$ is also small. We exploit these findings to reduce
the dimensionality of the problem: we can 
discount the variabilities in $T_u$, $T_l$, and $K_w$ by fixing 
them at their respective nominal values. 

\begin{figure}[htbp]
 \begin{center}
  \includegraphics[width=0.48\textwidth]{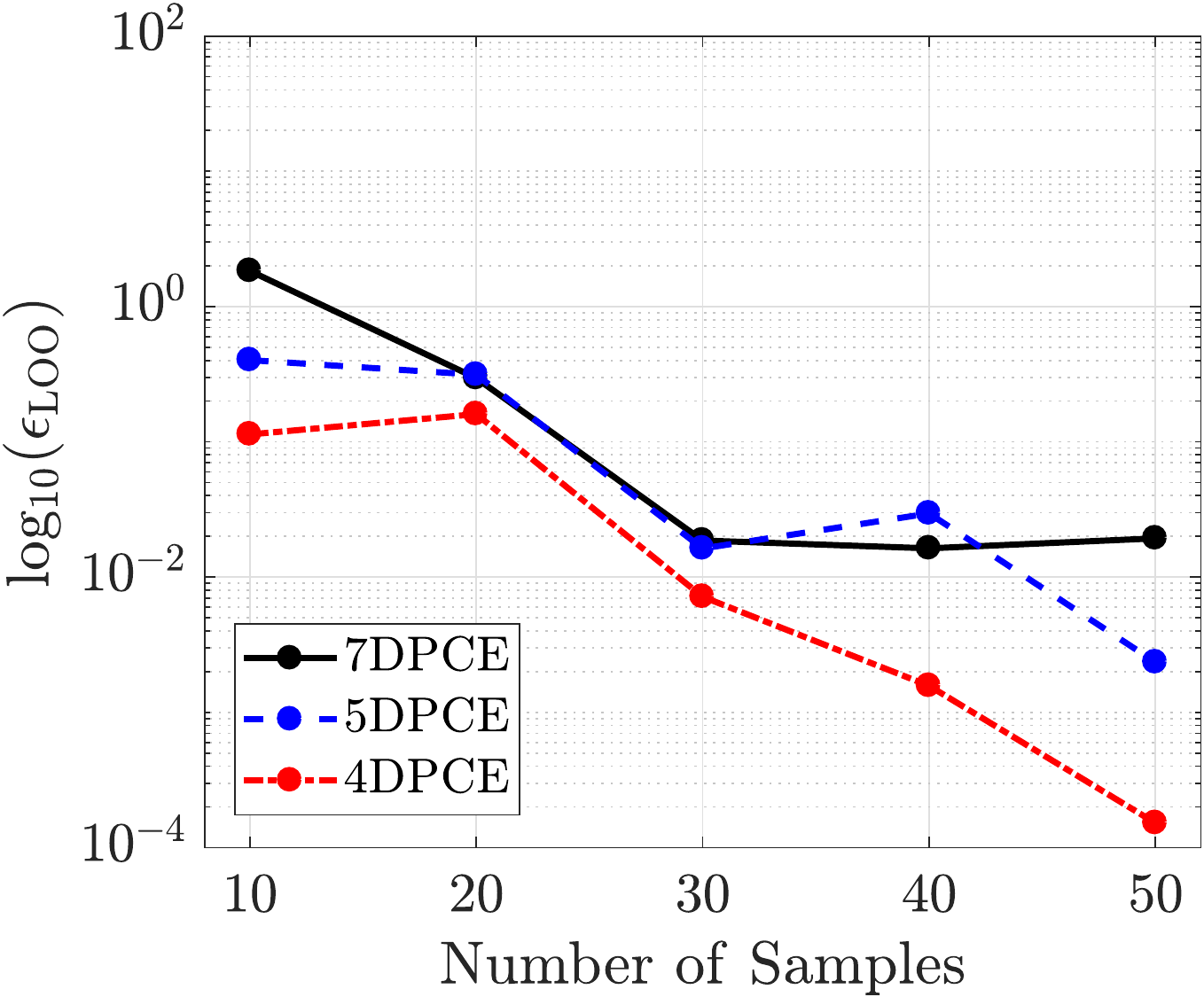}
  \includegraphics[width=0.48\textwidth]{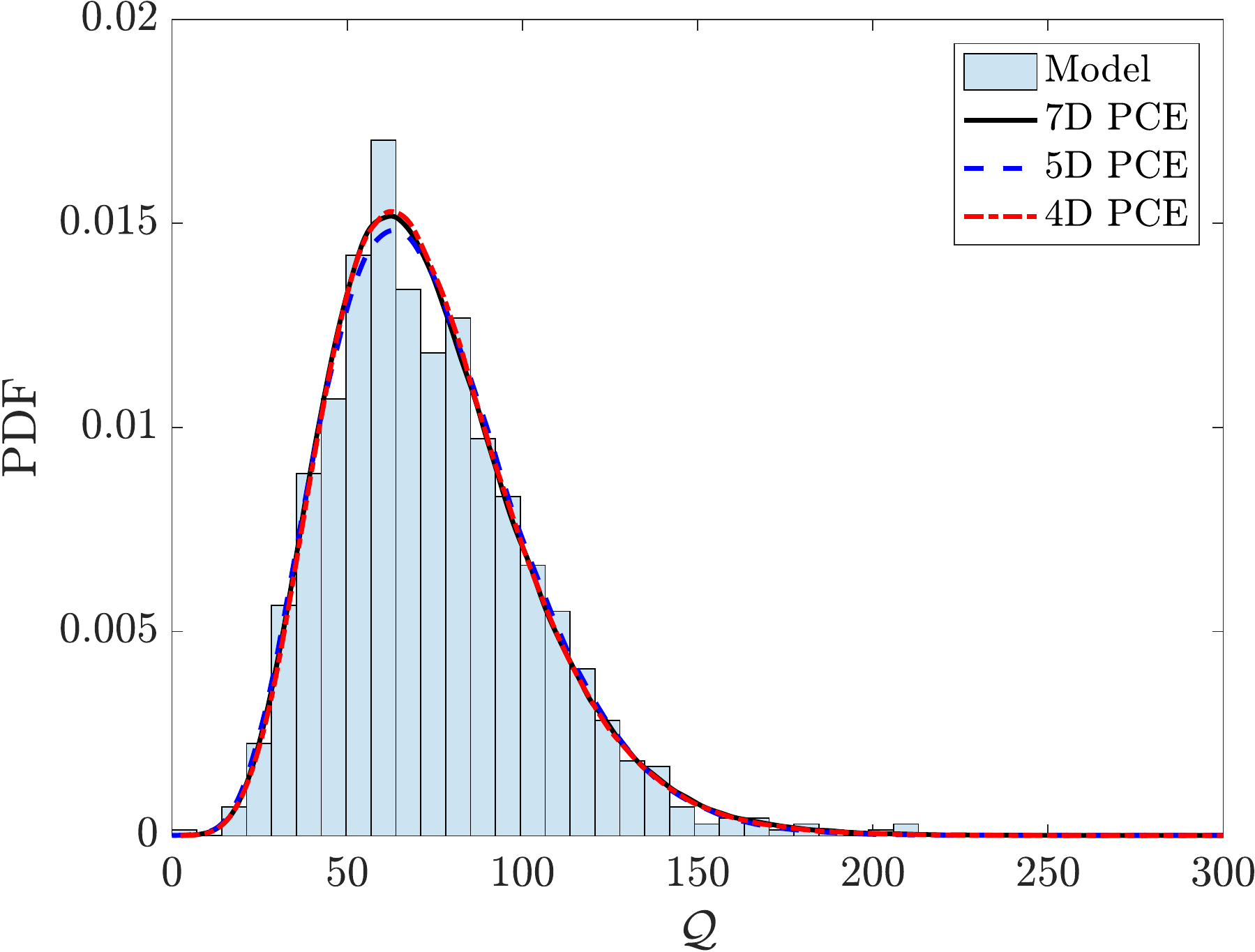}
\caption{Left: A comparison of order of the leave-one-out-error 
($\epsilon_{\mbox{\tiny{LOO}}}$) as a function of number of regression samples
used for constructing the PCE in 4, 5, and 7 dimensions. Right: A comparison of
PDFs of the discharge, $\mathcal{Q}$, generated using 10$^{6}$ samples from
the marginal distributions of the uncertain parameters in each case.} 
\label{fig:conv_bore}
\end{center}
\end{figure}

Our goal as discussed earlier is to gain computational advantage by constructing
surrogates in a reduced input parameter space. To this end, we use LAR to 
construct PCEs
in 5D and 4D spaces by fixing $\{T_u,T_l\}$ in the former and additionally
fixing $K_w$ in the latter at their respective mean values. In
Figure~\ref{fig:conv_bore}~(left), we compare convergence of PCEs constructed in the
full space (7D) with those constructed in the two reduced spaces (4D and 5D)
using $\epsilon_{\mbox{\tiny{LOO}}}$ (Eq.~\ref{eq:loo}). 

As expected, it is observed that the PCE constructed in the 4D space converges
at a much faster rate. For instance, if a PCE with $\mathcal{O}(10^{-4})$
accuracy is sought, we need function evaluations at only about 50 sample points
in the 4D parameter space whereas the number of samples needed in the full 7D
space seems much higher. Latin hypercube sampling (LHS) was used in
each case. It must be pointed out that the error in Figure~\ref{fig:conv_bore}~(left)
is not expected to decrease monotonically with the increase in sample size
owing to the penalty term in the regularized optimization problem in Eq.~\ref{eq:reg}.

As discussed earlier in this section, the reduced-space PCE's are verified for
predictive accuracy in a least-squares sense and a probabilistic sense.
Estimates for $\epsilon_{\mbox{\tiny{L-2}}}$ based on 50 samples in the
validation test suite were
found to be 0.0551 and 0.0112 for the 4D and 5D PCE's, respectively. In other
words, the 4D PCE is accurate within 5.52$\%$ and the 5D PCE is accurate within
1.12$\%$ of predictions based on the borehole function. Note that $\epsilon_\text{\tiny{LOO}}$
however, is lower in the case of 4D PCE (Figure~\ref{fig:conv_bore}).
This illustrates the
trade-off between accuracy and computational efficiency for the present 
problem. Generally, the required level of accuracy is problem dependent. 
The present framework allows for moving towards higher fidelity 
reduced-space surrogates based on the ranking of the parameter sensitivities. 
 

Figure~\ref{fig:conv_bore}~(right) illustrates a comparison of the PDFs
of the
discharge, $\mathcal{Q}$ obtained by propagating 10$^6$ random
samples through the 7D PCE in the original input parameter domain as well as
the reduced-space PCEs constructed in 4 and 5 dimensions. A normalized histogram
plot using 1000 model evaluations in the validation test suite is also included.
It is evident from this plot that the PDFs agree quite favorably with each
other as well as the original model-based histogram with respect to the modal estimate
as well as the uncertainty associated with $\mathcal{Q}$. 
Consequently, it can be said that the reduced-space PCE is verified in a
probabilistic sense. In other words, the mode as well as the uncertainty in the
output is reliably captured and predicted by the reduced-space PCE. 

\subsection{Semilinear elliptic PDE with random source term}

\begin{figure}[htbp]
\begin{center}
\begin{minipage}[htbp]{.25\linewidth}
\vspace{0pt}
\hspace{-25mm}
\begin{tabular}{cl}
\toprule
$\theta_i$ & \textbf{Distribution} \\ 
\bottomrule
$\kappa$ & $\mathcal{U}$[0.05,0.1] \\
$c$ & $\mathcal{U}$[1.0,2.0] \\
$\alpha_i$ & $\mathcal{U}$[0.0,4.0] \\
\bottomrule
\end{tabular}
\end{minipage}
\hspace{-20mm}
\begin{minipage}[htbp]{.25\linewidth}
\vspace{0pt}
\includegraphics[width=3.0in]{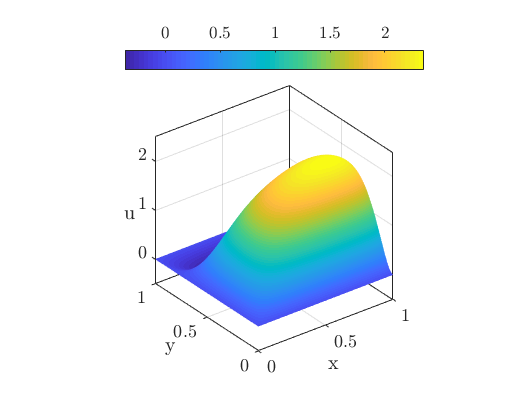}
\end{minipage}%
\end{center} 
\caption{Left: Table providing distributions of the individual
uncertain parameters in~\eqref{eq:semilinear}. Right: Solution
of the 2D semilinear elliptic PDE~\eqref{eq:semilinear} using
$\kappa$~=~0.075, $c$~=~1.5, and $\alpha_i$~=~4.0}
\label{fig:elliptic}
\end{figure}

We consider the following semilinear elliptic PDE: 
\begin{equation}\label{eq:semilinear}
\begin{aligned}
-\kappa \Delta u + c u^3 &= q \quad \text{in } \Omega,\\
 u &= 0 \quad \text{on } \partial \Omega.
\end{aligned}
\end{equation}
Here $\Omega = (0, 1)\times(0,1)$, 
$u$ is the state variable, and $\kappa$ and $c$ are coefficients of the diffusion term
and the nonlinear term in the above equation, respectively. 
We consider uncertainties in $\kappa$, $c$, and the source term. 
The right hand side function $q$ is defined by 
\be
q(x, y) = 
\sum\limits_{i=1}^{N=8}\alpha_i\sin\left(\frac{i\pi x}{8}\right)
                               \cos\left(\frac{i\pi y}{8}\right),
\label{eq:source}
\ee
where $\alpha_i$, $i = 1, \ldots, 8$ are random coefficients.
Hence, $u~=~u(\vec{\theta})$, where 
\[\vec{\theta} = 
\begin{bmatrix} \kappa & c & \alpha_1 & \alpha_2 & \cdots & \alpha_{8}
\end{bmatrix}^T
\] 
is the vector
of uncertain parameters. Distributions of the uncertain
input parameters are tabulated in Figure~\ref{fig:elliptic}~(left).
The solution of~\eqref{eq:semilinear} for
a fixed set of values of the uncertain parameters is also 
illustrated.

\begin{figure}[htbp]
 \begin{center}
  \includegraphics[width=0.46\textwidth]{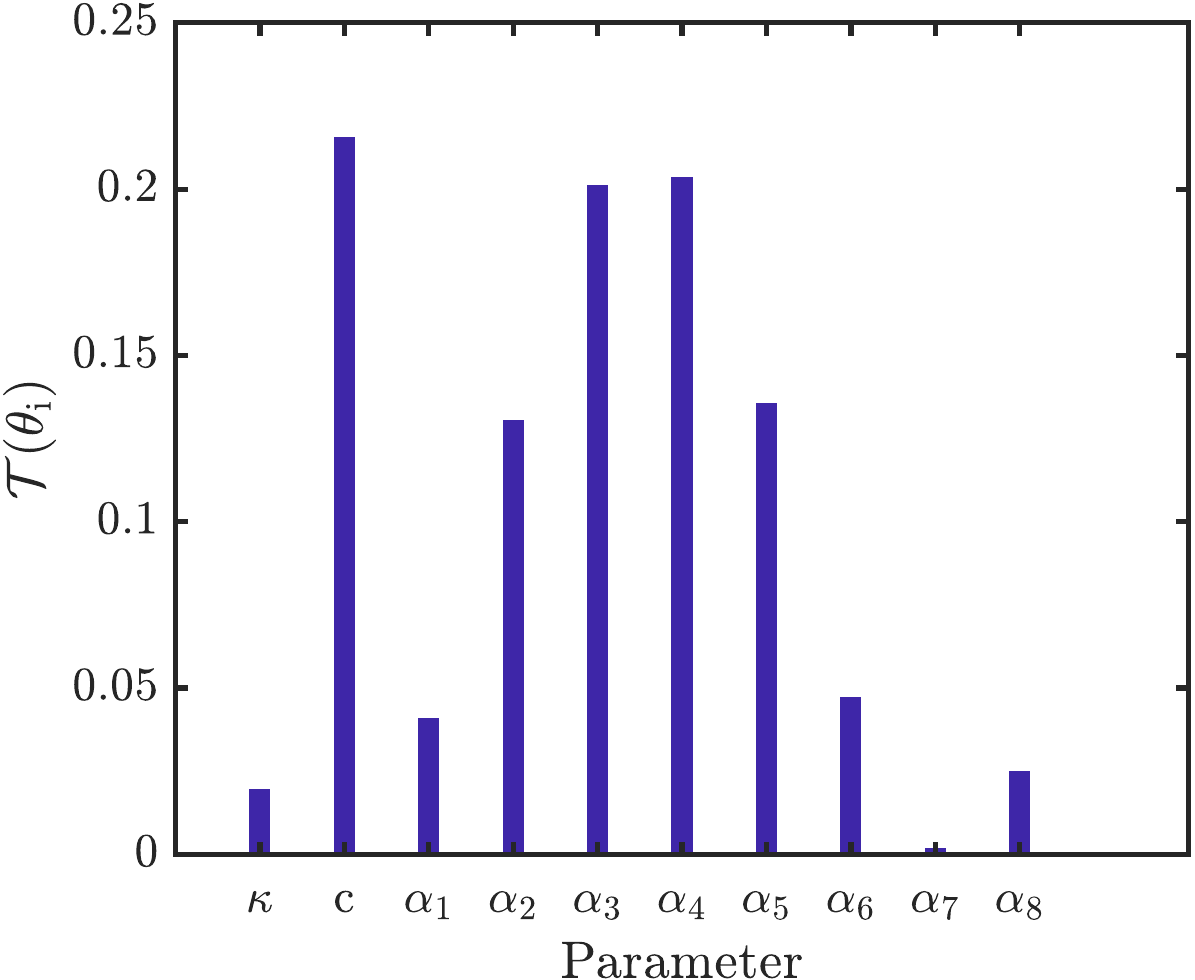}
  \includegraphics[width=0.48\textwidth]{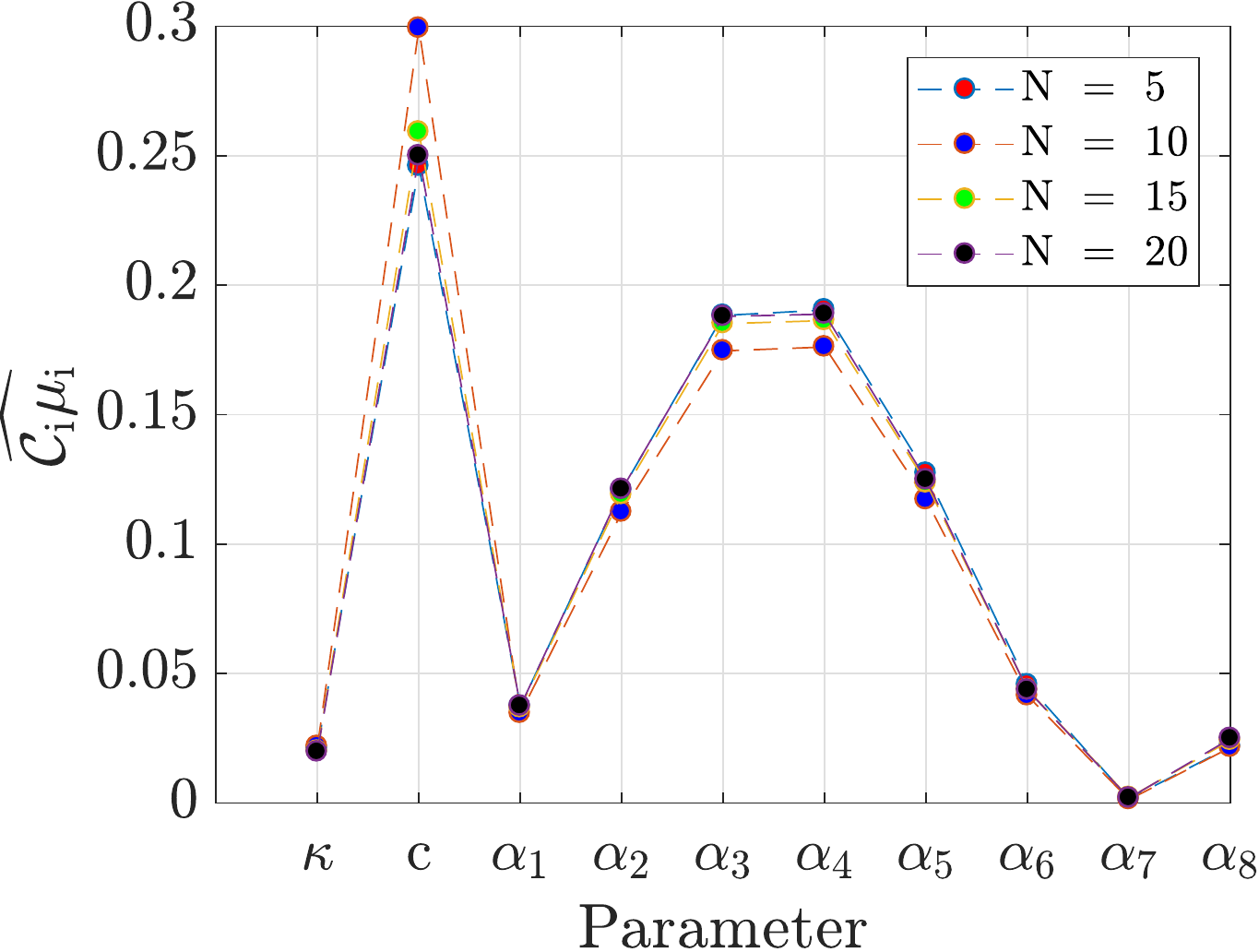}
\caption{
Left: Sobol' total sensitivity index, $\mathcal{T}(\theta_i)$ for uncertain parameters in the 
semilinear elliptic PDE~\eqref{eq:semilinear}. Right: 
Estimates of the screening metric ($\widehat{\mathcal{C}_i\mu_i}$) for each uncertain parameter,
obtained using $N$ = 5, 10, 15, and 20 samples in the full parameter space.}
\label{fig:sense_elliptic}
\end{center}
\end{figure}

\begin{figure}[htbp]
 \begin{center}
  \includegraphics[width=0.48\textwidth]{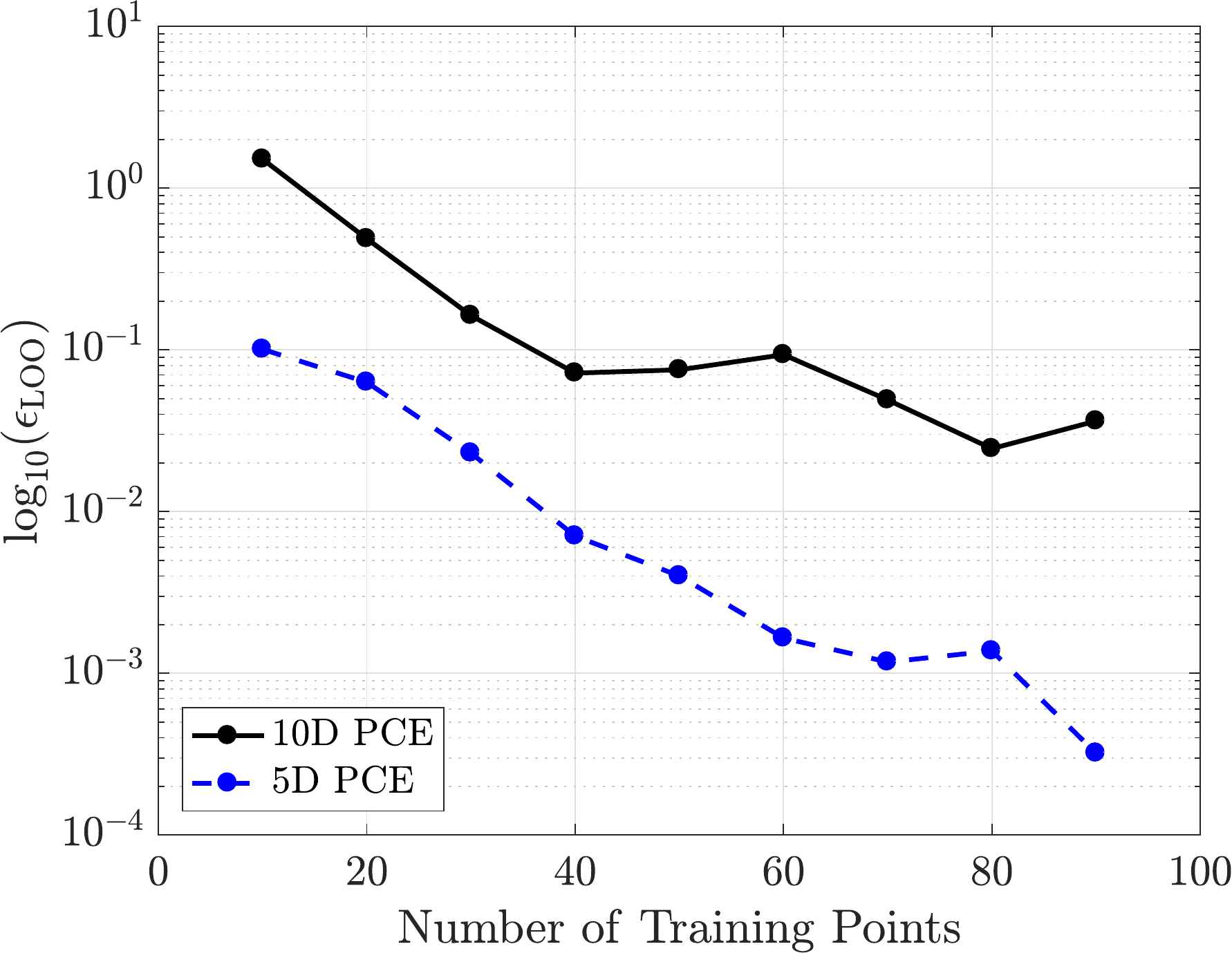}
  \includegraphics[width=0.48\textwidth]{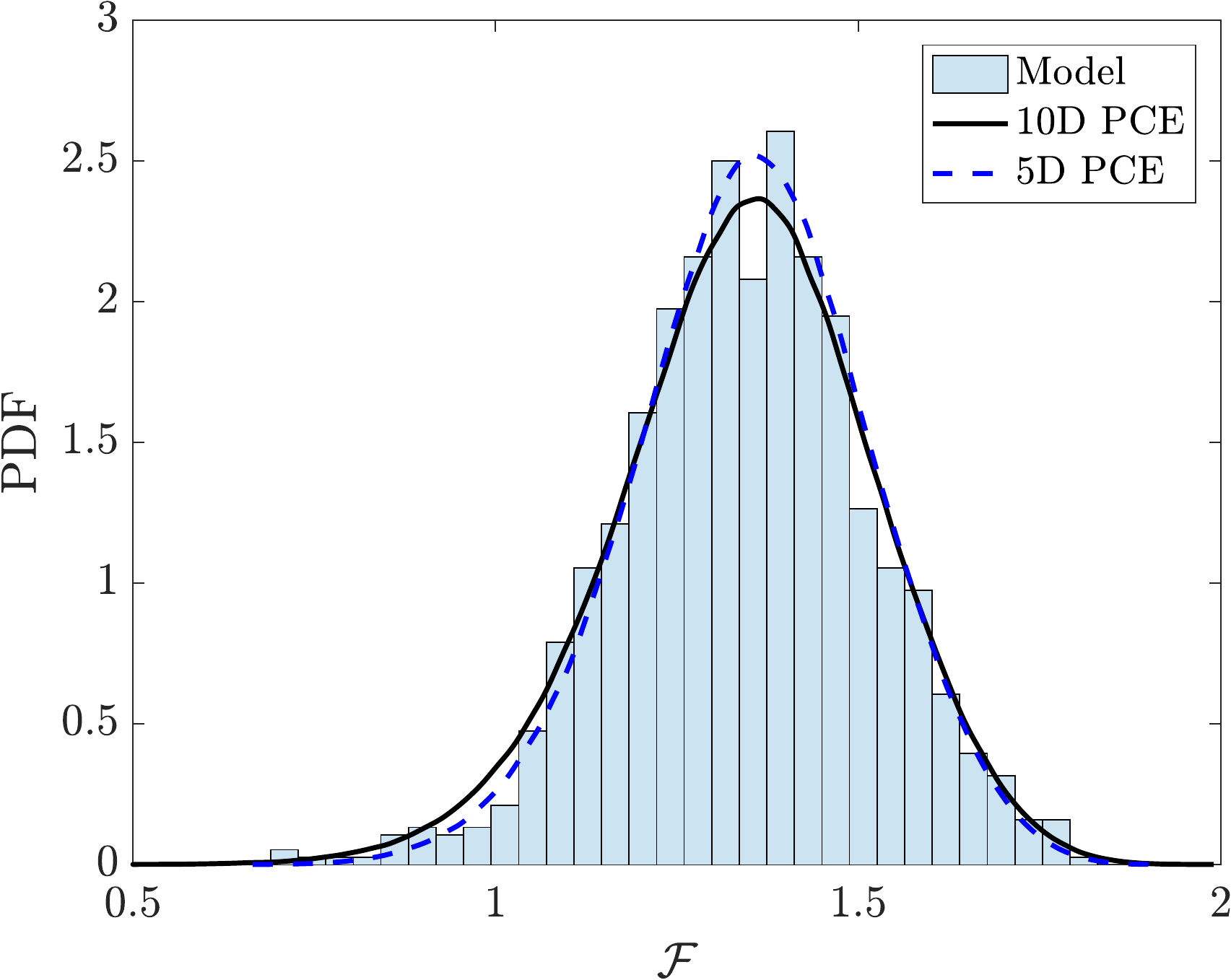}
\caption{Left: Logarithm of $\epsilon_{\mbox{\tiny{LOO}}}$ is plotted against sample size for 
PCEs constructed in 10 and 5 dimensions to compare their convergence characteristics. 
Right: PDF of the QoI, $\mathcal{F}(\vec\theta)$ in~\eqref{eq:semilinear} is
plotted using the full-surrogate and the reduced-space PC surrogate.}
\label{fig:conv_elliptic}
\end{center}
\end{figure}

We aim to construct a reduced-space surrogate for the following QoI:
\be
\mathcal{F}(\vec\theta) = \frac{1}{|D|} \int_D u(\vec{x}; \vec\theta) \, d\vec{x}, 
\label{eq:qoi}
\ee
where $D$ is the region $[2/5, 3/5] \times [2/5, 3/5] \subset \Omega$, 
and $|D|$ denotes the area of $D$. 
While this model is considerably more complex than the previous numerical
examples, it can still be solved efficiently.
The equation was discretized using finite differences, and Newton's method
was used to solve the resulting nonlinear system on a $100 \times 100$ 2D
cartesian grid.
We computed converged estimates of the Sobol
total-effect index $\mathcal{T}(\theta_i)$, reported 
in Figure~\ref{fig:sense_elliptic}~(left) using FSS predictions at 10$^6$
MC samples in the input space. The FSS was 
constructed using model predictions at 500 training points in the 10-dimensional
input space. Corresponding value of $\epsilon_\text{\tiny{LOO}}$ was 
found to be 9.729$\times 10^{-4}$. 
Sensitivity predictions based on the screening metric, $\widehat{\mathcal{C}_i\mu_i}$,
plotted in Figure~\ref{fig:sense_elliptic}~(right), are found to
be in close agreement with $\mathcal{T}(\theta_i)$, even for the case when $N$ = 5. As $N$
is increased from 5 to 20, estimates of the screening metric are observed to converge.
Based on the trends observed in Figure~\ref{fig:sense_elliptic}, it can be said that
the uncertainty in the QoI in \eqref{eq:qoi} is largely dependent on $c$, 
$\alpha_2$, $\alpha_3$, $\alpha_4$, and $\alpha_5$. These observations underscore the
potential for computational gains by constructing an RSS in the 5D parameter space. We 
illustrate the comparison of convergence characteristics of the PCEs constructed in the
full parameter space (10D) and the reduced space (5D) in Figure~\ref{fig:conv_elliptic} (left). 
As expected, the RSS converges considerably faster. Using model evaluations at 90
sample points, $\epsilon_{\mbox{\tiny{LOO}}}$ is found to be two orders of magnitude
smaller than that in the case of full-surrogate ($\mathcal{O}(10^{-4}$) versus
$\mathcal{O}(10^{-2}$)). 
Consequently, the computational effort for constructing the RSS in the present test problem
is expected to be much smaller. 

Once again, we verify the accuracy of the RSS by estimating $\epsilon_{\mbox{\tiny{L-2}}}$
using model evaluations at 1000 independent MC samples in the 10D parameter
space. The RSS was found to be accurate
within 5$\%$. In order to bolster confidence in the RSS, we compare PDFs of the QoI
as well as a normalized histogram plot based on sparse model evaluations in the  
validation test-suite, in 
Figure~\ref{fig:conv_elliptic}~(right). While the two PDFs are in favorable agreement,
the modal estimate and the spread in the QoI based on the histogram is also captured
by them. Hence, the RSS could be used with a reasonable degree of confidence to
quantify the uncertainty in $\mathcal{F}(\vec\theta)$ thereby leading to
a computational advantage in this case.

\bigskip
\bigskip

\section{Application: H$_2$/O$_2$ Reaction Kinetics}
\label{sec:app}

%
%

The proposed framework in section~\ref{sec:method} is implemented to 
the H$_2$/O$_2$ reaction mechanism provided in~\cite{Yetter:1991}.
The H$_2$/O$_2$ reaction is gaining a lot of attention as a potential
source of clean energy for applications such as 
transportation~\cite{Das:1996}. 
The mechanism comprises of 19 reactions including chain reactions,
dissociation/recombination reactions, and formation and consumption
of intermediate species as provided below in Table~\ref{tab:kinetics}.

\begin{table}[htbp]
\renewcommand*{\arraystretch}{1.2}
\begin{center}
\begin{tabular}{llll}
\toprule
Reaction \#     & Reaction &&\\
\bottomrule
$\mathcal{R}_1$ & H + O$_2$          & $\rightleftharpoons$ & O + OH \\
$\mathcal{R}_2$ & O + H$_2$          & $\rightleftharpoons$ & H + OH \\
$\mathcal{R}_3$ & H$_2$ + OH         & $\rightleftharpoons$ & H$_2$O + H \\
$\mathcal{R}_4$ & OH + OH            & $\rightleftharpoons$ & O + H$_2$O \\
$\mathcal{R}_5$ & H$_2$ + M          & $\rightleftharpoons$ & H + H + M \\
$\mathcal{R}_6$ & O + O + M          & $\rightleftharpoons$ & O$_2$ + M \\
$\mathcal{R}_7$ & O + H + M          & $\rightleftharpoons$ & OH + M \\
$\mathcal{R}_8$ & H + OH +M          & $\rightleftharpoons$ & H$_2$O + M \\
$\mathcal{R}_9$ & H + O$_2$ + M      & $\rightleftharpoons$ & HO$_2$ + M \\
$\mathcal{R}_{10}$ & HO$_2$ + H      & $\rightleftharpoons$ & H$_2$ + O$_2$ \\
$\mathcal{R}_{11}$ & HO$_2$ + H      & $\rightleftharpoons$ & OH + OH \\
$\mathcal{R}_{12}$ & HO$_2$ + O      & $\rightleftharpoons$ & O$_2$ + OH \\
$\mathcal{R}_{13}$ & HO$_2$ + OH     & $\rightleftharpoons$ & H$_2$O + O$_2$ \\
$\mathcal{R}_{14}$ & HO$_2$ + HO$_2$ & $\rightleftharpoons$ & H$_2$O$_2$ + O$_2$ \\
$\mathcal{R}_{15}$ & H$_2$O$_2$ + M  & $\rightleftharpoons$ & OH + OH + M \\
$\mathcal{R}_{16}$ & H$_2$O$_2$ + H  & $\rightleftharpoons$ & H$_2$O + OH \\
$\mathcal{R}_{17}$ & H$_2$O$_2$ + H  & $\rightleftharpoons$ & HO$_2$ + H$_2$ \\
$\mathcal{R}_{18}$ & H$_2$O$_2$ + O  & $\rightleftharpoons$ & OH + HO$_2$ \\
$\mathcal{R}_{19}$ & H$_2$O$_2$ + OH & $\rightleftharpoons$ & HO$_2$ + H$_2$O \\
\bottomrule
\end{tabular}
\end{center}
\caption{Reaction mechanism for H$_2$/O$_2$ from~\cite{Yetter:1991}}.
\label{tab:kinetics}
\end{table}

The reaction rate for the $i^{th}$ reaction as a function of temperature
is given as follows:
\be
k_i(T) = A_iT^{n_i}\exp(-E_{a,i}/RT), 
\label{eq:rate}
\ee
where $A_i$ is the pre-exponent, $n_i$ is the index of $T$, $E_{a,i}$ is the
activation energy corresponding to the $i^{th}$ reaction, and $R$ is the
universal gas constant. 
The TChem~\cite{Safta:2011} software package is used to model homogeneous
ignition at constant pressure for a range of initial conditions for the
fuel-oxidizer mixture. During the simulation, the fuel-oxidizer mixture goes 
through a radical build-up phase followed by a sharp increase in temperature
as heat is released during the thermal runaway. We focus on quantifying the 
uncertainty in the \emph{ignition delay} due to uncertainty associated 
with the pre-exponent, $A_i$, for each reaction. The ignition delay 
is defined as the inflection point on the temperature profile during the thermal
runaway. The total number of uncertain parameters in the
present case is 19.  The $A_i$'s are considered to be uniformly distributed in
the interval: $[0.9A_i^\ast, 1.1A_i^\ast]$; $A_i^\ast$ being the nominal
estimate corresponding to the $i^{th}$ reaction.
The set of nominal values used in the computations, for parameters 
in~\eqref{eq:rate} are provided
in~\cite{Yetter:1991}. 

While the dimensionality of the problem is relatively moderate,
constructing a surrogate in the 19-dimensional parameter space could still be
expensive. Hence, we explore the possibility of constructing a
reduced-space surrogate (RSS) using the framework presented 
in section~\ref{sec:method}. 
In the present study, we focus on two scenarios: fuel(H$_2$)-rich, and
fuel(H$_2$)-lean. Consider the global reaction:
\be
2\text{H}_2 + \text{O}_2 \rightarrow 2\text{H}_2\text{O}
\label{eq:global}
\ee 
The equivalence ratio $\phi$ is defined as follows:
\be
\phi = \frac{(M_{\text{H}_2}/M_{\text{O}_2})_\text{obs}}{(M_{\text{H}_2}/M_{\text{O}_2})_\text{st}}
\label{eq:phi}
\ee
The numerator in the right-hand-side represents the observed (obs) fuel-oxygen
mass ratio at a given condition and the denominator represents the
stoichiometric (st) ratio of the same quantity. Hence, $\phi$ = 1 at
stoichiometric conditions. The equivalence ratio can be altered by changing the
amount of O$_2$ in the mixture. In the case of a lean
mixture,~\eqref{eq:global} can be written as follows:
\be
2\text{H}_2 + \alpha\text{O}_2 \rightarrow 2\text{H}_2\text{O} + (\alpha-1)\text{O}_2 
\hspace{3mm} (\alpha>1)
\label{eq:lean}
\ee 
Similarly, for the case when the mixture if fuel rich,~\eqref{eq:global} is modified
as follows:
\be
2\text{H}_2 + \alpha\text{O}_2 \rightarrow 2\alpha\text{H}_2\text{O} + 2(1-\alpha)\text{H}_2
\hspace{3mm} (\alpha<1)
\label{eq:rich}
\ee 
Eqs.~\eqref{eq:lean} and~\eqref{eq:rich} can be generalized as follows:
\be
2\text{H}_2 + \alpha\text{O}_2 \rightarrow 2\min(1,\alpha)\text{H}_2\text{O} + 
\max(\alpha-1,0)\text{O}_2 + \max(0,2-2\alpha)\text{H}_2
\label{eq:gen}
\ee 
From the above set of chemical equations, the relationship between $\phi$
and $\alpha$ can be easily obtained as $\phi~=~\frac{1}{\alpha}$.
Since $\phi>1$ corresponds to a rich mixture, and $\phi<1$ corresponds to a
lean mixture, we consider $\phi$ = 2.0 and 0.5 to investigate the two scenarios
respectively. 

We apply the parameter screening algorithm with the following
parameters: $\tau_\text{screen}$, $s_\text{min}$,
$s_\text{max}$, $\beta$ are fixed at 0.2, 3, 10, and 1.0 respectively for both cases.
Additionally, the value of $\tau$ is considered to be 1.0$\times 10^{-17}$ and
5.0$\times 10^{-17}$ in the rich and lean case respectively. Such a small value
of $\tau$ for this application is a consequence of the nature of convergence exhibited
by the sensitivity measures. Moreover, the screening procedure is carried out
for atleast $s_\text{min}$ number of iterations in order to bolster our confidence
in the estimates. 

Following the steps outlined in the flow-diagram in Figure~\ref{fig:flow}, model
evaluations are initially generated at $n_1$ = 5 samples. The evaluations are
used to construct a regression-based surrogate in the full-space. As
expected, the surrogate is found to be highly inaccurate. Moreover, unlike the
test problems in section~\ref{sec:examples}, we do not estimate the Sobol
total-effect sensitivity indices in the interest of following the overall framework
closely. Hence, we proceed to the screening step to estimate the screening metric
for the uncertain pre-exponents, $A_i$'s. Results are plotted below
in Figure~\ref{fig:sense_kinetics}~(top row) for both cases. Furthermore, we illustrate
the decay in the value of $\Delta\mu_s$ with iterations in 
Figure~\ref{fig:sense_kinetics}~(bottom row). 

\begin{figure}[htbp]
 \begin{center}
  \includegraphics[width=0.48\textwidth]{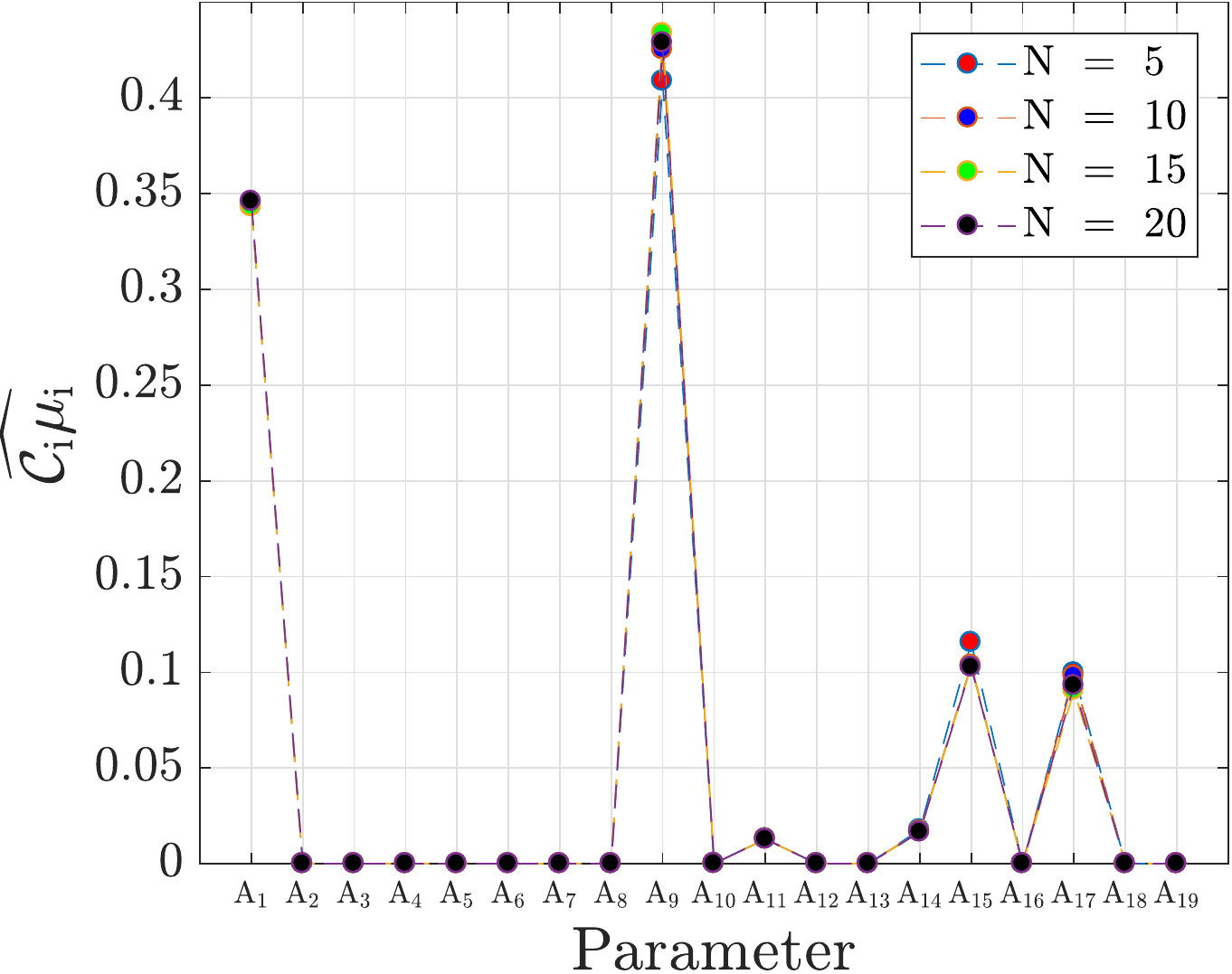}
  \includegraphics[width=0.48\textwidth]{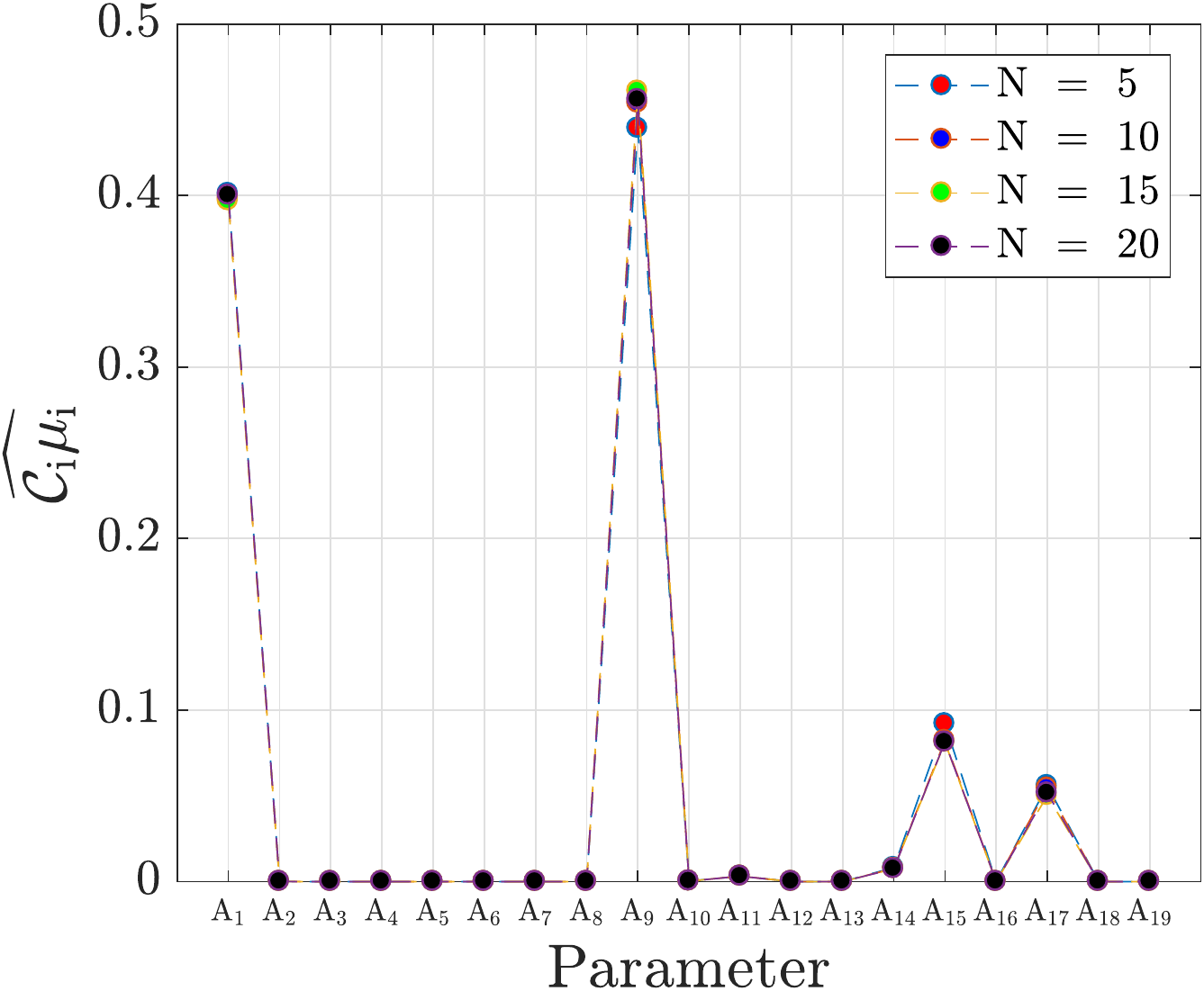}
  \\ \vspace{2mm}
  \includegraphics[width=0.48\textwidth]{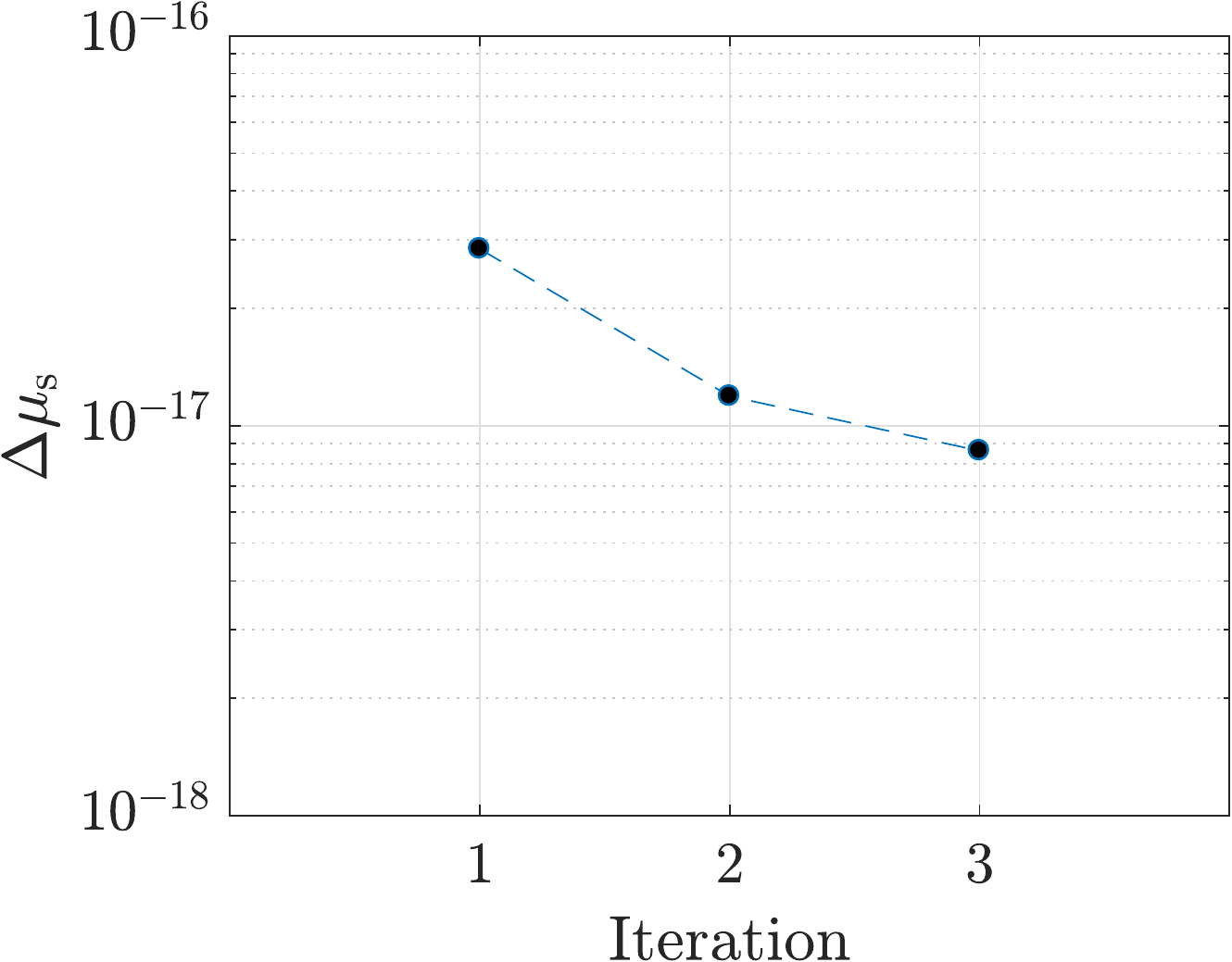}
  \includegraphics[width=0.48\textwidth]{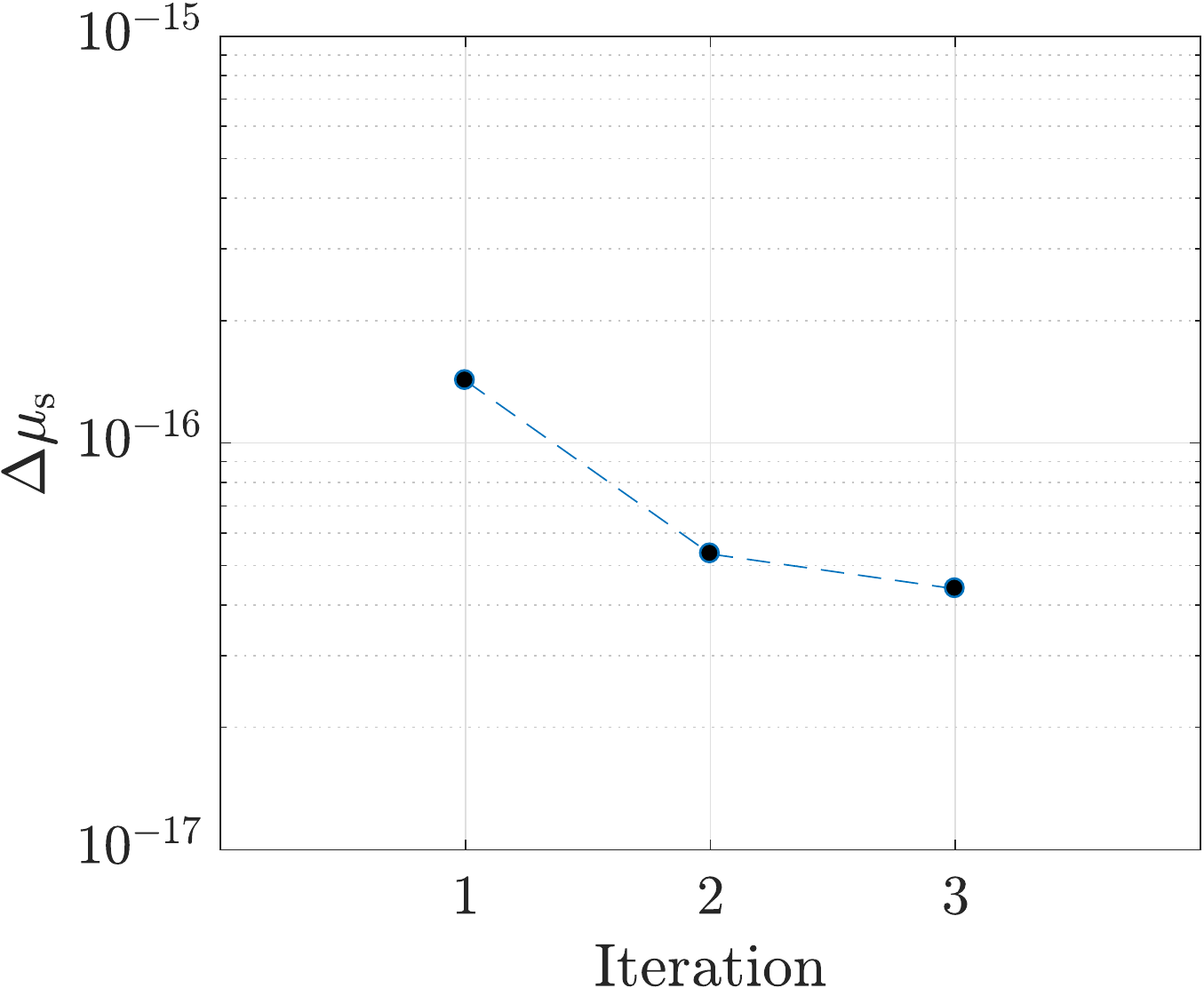}
\caption{Top: Estimates of $\widehat{\mathcal{C}_i\mu_i}$ for $A_i$'s in the case
of fuel-rich mixture~(left) and fuel-lean mixture~(right). Bottom: The value of
$\Delta\mu_s$ during three iterations within the screening step are plotted for
the case of fuel-rich mixture~(left) and fuel-lean mixture~(right).}
\label{fig:sense_kinetics}
\end{center}
\end{figure}
The screening metric estimates in the above plots are observed to converge with
only a few samples (5--10). Moreover, out of the 19 uncertain pre-exponents,
only $A_1$, $A_9$, $A_{15}$, and $A_{17}$ seem to be important in the fuel-rich
case, whereas, only $A_1$, $A_9$, and $A_{15}$ seem important in the fuel-lean
case, based on the value of $\tau_\text{screen}$. These observations are indicative
of the potential for significant reduction in the dimensionality of this problem. 
A reduced-space
surrogate constructed using the proposed framework could thus lead to large computational
gains. The decay of $\Delta\mu_s$ with iterations is expected and builds our confidence
in the screening procedure in both cases.

\begin{figure}[htbp]
 \begin{center}
  \includegraphics[width=0.48\textwidth]{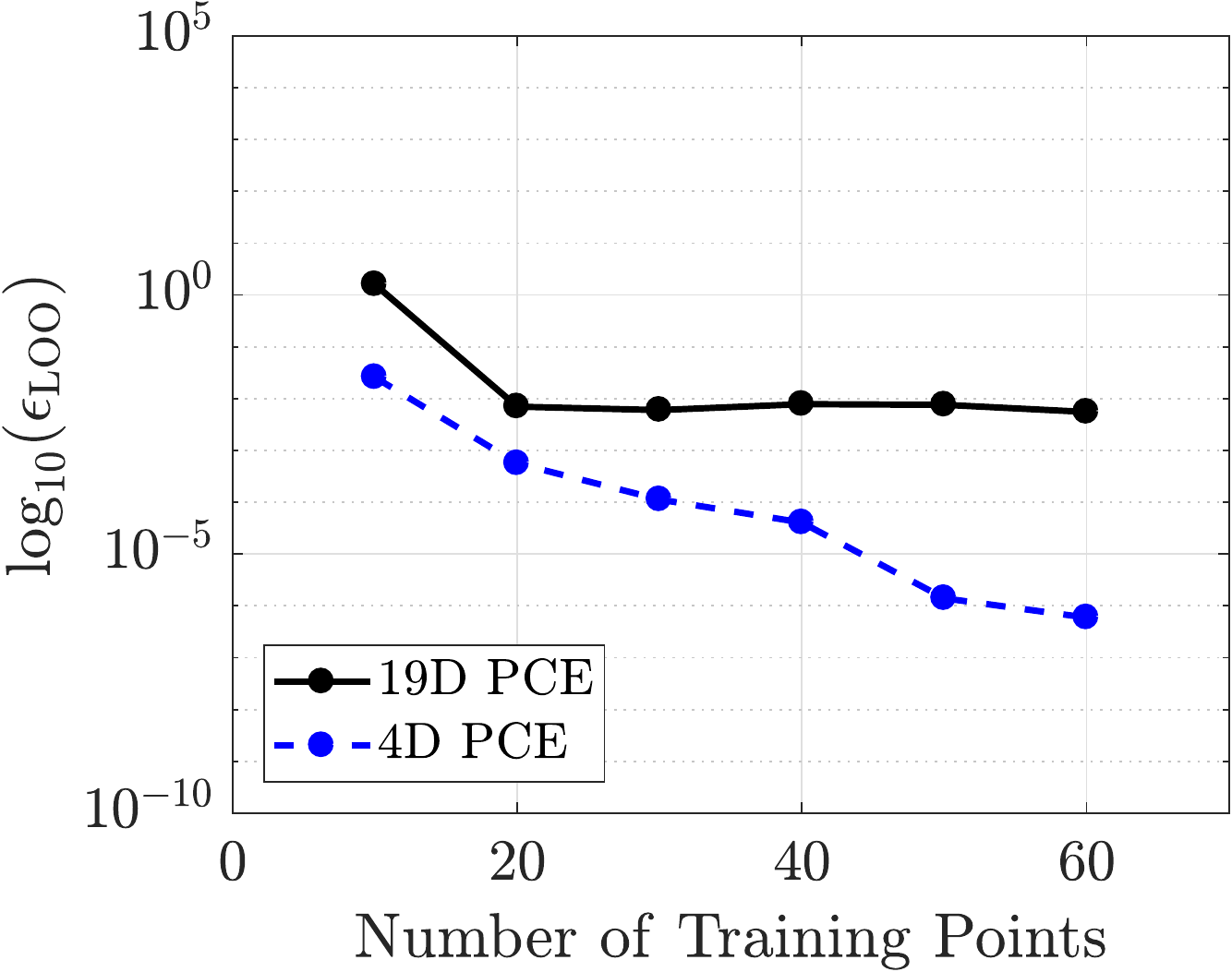}
\caption{A semi-log plot of $\epsilon_\text{\tiny LOO}$ as a function of
number of model evaluations in the full-space (19D) and the reduced-space (4D)
for the fuel (H$_2$)-rich case i.e. $\phi$ = 2.0.}
\label{fig:err_samples_kinetics}
\end{center}
\end{figure}

A reduced-space surrogate (RSS) was constructed in 4D for the fuel-rich case,
and in 3D for
the fuel-lean case. Figure~\ref{fig:err_samples_kinetics} illustrates a
comparison of convergence characteristics for the PCEs constructed in the
full-space and the reduced-space for the fuel-rich case. Note that the
plot is generated using the implementation of least angle regression (LAR)
for sparse PCEs in UQLab.  
The leave-one-out cross validation error is observed to drop initially
and plateau with the increase in training points for the 19-dimensional
PCE. However, in the case of 4-dimensional PCE, the error exhibits a
monotonic behavior and is found to be smaller than $\mathcal{O}(10^{-5})$
at 60 training points. Clearly, the RSS shows a much faster rate of convergence.
Similar trends (not included) were observed in the fuel-lean case. 

\begin{figure}[htbp]
 \begin{center}
  \includegraphics[width=0.48\textwidth]{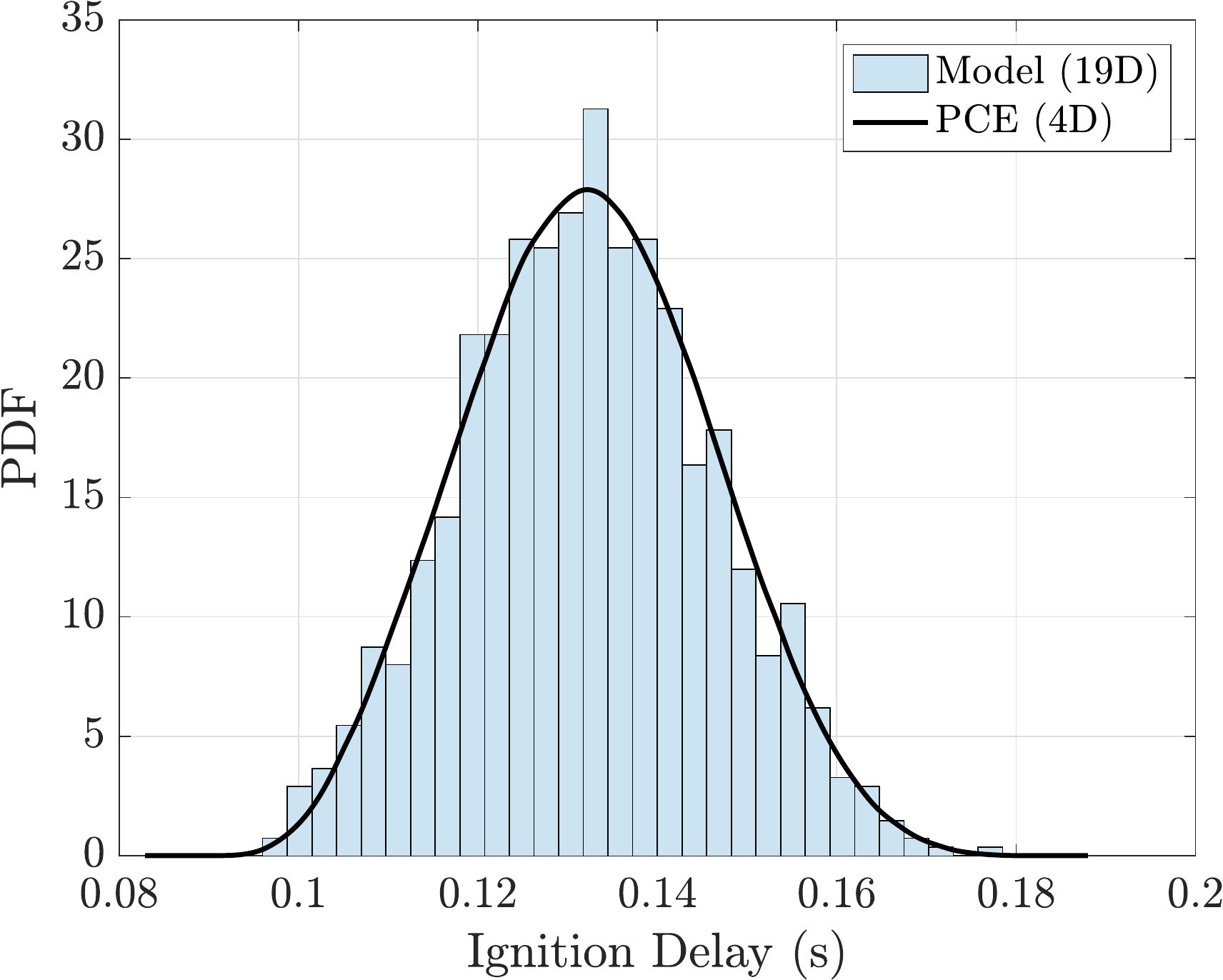}
  \includegraphics[width=0.48\textwidth]{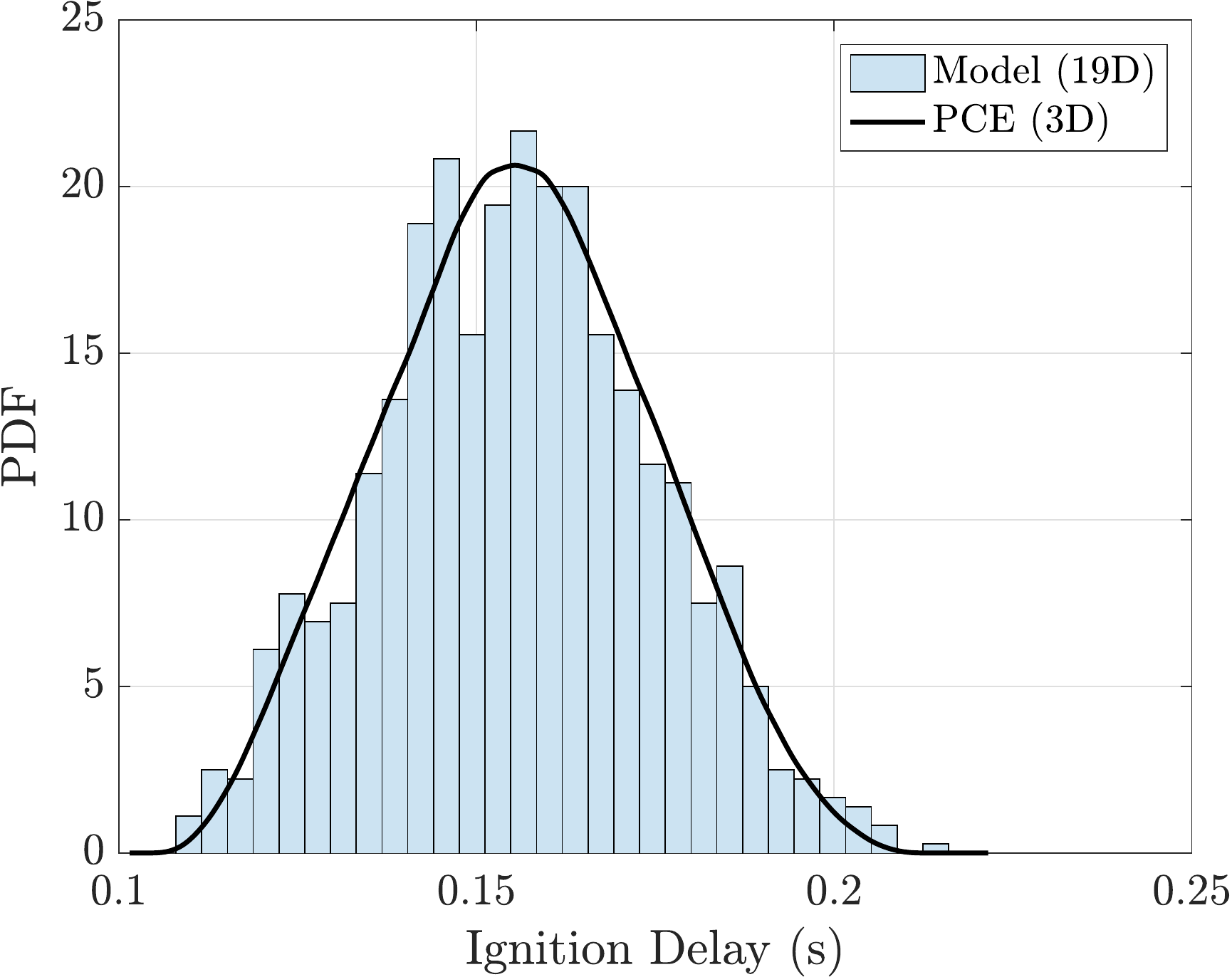}
\caption{A normalized histogram based on model evaluations at 1000 samples is plotted
along with a PDF of ignition delay for the fuel-rich case (left) and the fuel-lean
case (right).}
\label{fig:pdf_kinetics}
\end{center}
\end{figure}

Based on $\epsilon_{\mbox{\tiny{L-2}}}$ estimates using the 
cross-validation set, the RSS was found to be accurate within 1.8$\%$ in the
fuel-rich case, and
within 3.1$\%$ in the fuel-lean case. Model evaluations at 1000 samples in the test suite
are further used to plot a normalized histogram of the ignition time in 
Figure~\ref{fig:pdf_kinetics}. To verify the accuracy of the RSS in a 
probabilistic-sense, we compare the histogram plot with a PDF of ignition time
using surrogate predictions at 10$^{6}$ samples in the reduced space in both cases. 
Clearly, the RSS captures the spread as well as the modal estimate of
the ignition delay in both scenarios. Hence, the proposed framework 
has enabled significant dimension reduction and construction of an accurate
RSS for multiple scenarios pertaining to the H$_2$/O$_2$ reaction
mechanism.    

\bigskip
\bigskip

\section{Summary and Conclusion}
\label{sec:disc}


In this work, we have presented an efficient and practical approach for constructing
a reduced-space
surrogate for scientific and engineering applications. Dimension reduction is accomplished
by identifying uncertain parameters that contribute relatively less towards the uncertainty
in the quantity of interest. These parameters deemed as \textit{unimportant} are determined
using a screening metric~\eqref{eq:cmu} involving derivative-based sensitivity
measures. Initially, the metric is estimated
using model evaluations at a small set of samples in the parameter domain. These
estimates are refined by subsequent enrichment of the sample set during the screening
procedure presented in Algorithm~\ref{alg:screen}. The outcome of parameter screening is
assessed for the scope of dimension reduction. In a favorable scenario, a reduced-space
surrogate (RSS) is constructed. The RSS is tested for accuracy in a least-squares sense
as well as a probabilistic sense using a cross-validation test suite. In the proposed framework,
a surrogate in the full-space (FSS) is constructed in tandem with parameter screening using
the available set of model evaluations. Both, RSS and FSS are
constructed using regression-based sparse PCEs. Simultaneous construction
of the FSS ensures that the computational effort associated with the proposed framework
does not overshoot the effort required to construct the FSS directly.
Hence, the RSS is constructed only in situations where computational gains are
expected. 

Parameter screening methodology was implemented to low-to-moderate dimensional
test problems and an accurate RSS was constructed to demonstrate potential for
computational gains in each case. Furthermore, the overall framework was
implemented  to a relatively higher dimensional application involving kinetics
of the H$_2$/O$_2$ reaction mechanism.  Significant dimension reduction (19 
dimensions to 3 or 4 dimensions) was
accomplished for two different scenarios involving a fuel-rich and a fuel-lean
mixture. In both cases, the resulting RSS was able to capture the input-output
relationship as well as the uncertainty in the quantity of interest with
reasonable accuracy. Additional highlights of the proposed framework are as
follows:
\begin{enumerate}
\item Although PCEs were used in this work, the proposed framework is
agnostic to the choice of the surrogate model construction method.
\item Substantial computational gains are expected in situations involving
compute-intensive simulations even if the scope for dimension reduction is
small. 

\item Significant gains can be realized in situations where multiple surrogates need to be
constructed as illustrated in the kinetics application. Other possible scenarios may
include inverse problems involving parameter estimation in a Bayesian setting.

\item Dimension reduction based on the proposed methodology could help reduce the
effort required for model calibration wherein only the important parameters
are calibrated. 
\end{enumerate}

Based on the results presented for the test problems and the kinetics
application, the proposed framework seems quite promising in its
potential for identifying the unimportant model inputs. These
observations could be exploited to construct efficient model surrogates
in a reduced input space. However, it is important to remain cognizant
about the limitations of the framework as well. 
For instance, the quantity of interest is required to be differentiable with
respect to each parameter in the considered domain. This condition once
satisfied, enhances the accuracy of the PCE-based surrogates as well.  
Additionally, the proposed framework does not
account for the existence of possible correlations between the uncertain
inputs of the model. However, while the assumption of independent
inputs is not always justified, in many cases, correlations between
inputs are not well understood a priori, and assuming mutual independence
could be reasonable at least in initial screening using DGSMs. On the other
hand, if approximate correlations are known, we recommend using a Gaussian 
process or Kriging-based surrogate since it 
provides a means for incorporating the correlation between inputs.  
Implementation to applications involving strongly correlated parameters
could enhance the applicability of the proposed framework. 
We consider that to be a potential direction for future studies related
to this work.

\bigskip
\bigskip

\section*{Acknowledgment}

M. Vohra and S. Mahadevan gratefully acknowledge funding support from the
National Science Foundation (Grant No. 1404823, CDSE Program). 
C. Safta was supported by the U.S. Department of Energy, Office of Science,
Basic Energy Sciences, as part of the Computational Chemical Sciences Program.
M. Vohra would also like to sincerely thank Dr. Xun Huan at Sandia National Labs for
his guidance pertaining to the usage of TChem for the chemical kinetics application
in this work. Sandia National Laboratories is a multimission laboratory managed 
and operated by National Technology \& Engineering Solutions of Sandia, LLC, 
a wholly owned subsidiary of Honeywell International Inc., for the 
U.S. Department of Energy's National Nuclear Security Administration under contract DE-NA0003525. 
The views expressed in the article do not necessarily represent the
views of the U.S. Department Of Energy or the United States Government.
\bigskip
\bigskip

\bibliographystyle{unsrt}
\bibliography{REFER}

\begin{thebibliography}{10}

\bibitem{Xiu:2002}
D.~Xiu and G.E. Karniadakis.
\newblock The wiener--askey polynomial chaos for stochastic differential
  equations.
\newblock {\em SIAM journal on scientific computing}, 24(2):619--644, 2002.

\bibitem{Ghanem:2003}
R.G. Ghanem and P.D. Spanos.
\newblock {\em Stochastic finite elements: a spectral approach}.
\newblock Courier Corporation, 2003.

\bibitem{Olivier:2010}
O.~Le~Ma{\^\i}tre and O.M. Knio.
\newblock {\em Spectral methods for uncertainty quantification: with
  applications to computational fluid dynamics}.
\newblock Springer Science \& Business Media, 2010.

\bibitem{friedman93}
J.H. {Friedman}.
\newblock Fast {MARS}.
\newblock Technical Report 110, Laboratory for Computational Statistics,
  Department of Statistics, Stanford University, 1993.

\bibitem{Rasmussen:2004}
C.E. Rasmussen.
\newblock Gaussian processes in machine learning.
\newblock In {\em Advanced lectures on machine learning}, pages 63--71.
  Springer, 2004.

\bibitem{Stein:2012}
M.L. Stein.
\newblock {\em Interpolation of spatial data: some theory for kriging}.
\newblock Springer Science \& Business Media, 2012.

\bibitem{Sobol93}
I.M. {Sobol'}.
\newblock Sensitivity estimates for nonlinear mathematical models.
\newblock {\em Math. Mod. Comp. Exp.}, 1:407--414, 1993.

\bibitem{Sobol:2001}
I.M. Sobol'.
\newblock Global sensitivity indices for nonlinear mathematical models and
  their monte carlo estimates.
\newblock {\em Mathematics and computers in simulation}, 55(1):271--280, 2001.

\bibitem{Owen13}
A.B. {Owen}.
\newblock Better estimation of small {Sobol'} indices sensitivity indices.
\newblock {\em ACM Trans. Mod. Comput. Simul.}, 23:11--1:11--17, 2013.

\bibitem{SaltelliRattoAndresEtAl08}
A.~{Saltelli}, M.~{Ratto}, T.~{Andres}, F.~{Campolongo}, J.~{Cariboni},
  D.~{Gatelli}, M.~{Saisana}, and S.~{Tarantola}.
\newblock {\em Global sensitivity analysis: the primer}.
\newblock Wiley, 2008.

\bibitem{Sudret08}
B.~Sudret.
\newblock Global sensitivity analysis using polynomial chaos expansions.
\newblock {\em Reliability Engineering \& System Safety}, 93(7):964 -- 979,
  2008.

\bibitem{CrestauxLeMaitreMartinez09}
T.~Crestaux, O.P.~Le Maitre, and J.-M. Martinez.
\newblock Polynomial chaos expansion for sensitivity analysis.
\newblock {\em Reliability Engineering \& System Safety}, 94(7):1161 -- 1172,
  2009.
\newblock Special Issue on Sensitivity Analysis.

\bibitem{BlatmanSudret10}
G.~Blatman and B.~Sudret.
\newblock Efficient computation of global sensitivity indices using sparse
  polynomial chaos expansions.
\newblock {\em Reliability Engineering \& System Safety}, 95(11):1216--1229,
  2010.

\bibitem{HartAlexanderianGremaud17}
J.~Hart, A.~Alexanderian, and P.~Gremaud.
\newblock Efficient computation of sobol' indices for stochastic models.
\newblock {\em SIAM Journal on Scientific Computing}, to appear, 2017.

\bibitem{Sargsyan17}
K.~Sargsyan.
\newblock {\em Surrogate Models for Uncertainty Propagation and Sensitivity
  Analysis}.
\newblock Springer International Publishing, 2017.

\bibitem{AlexanderianWinokurSrajEtAl12}
A.~Alexanderian, J.~Winokur, I.~Sraj, A.~Srinivasan, M.~Iskandarani, W.C.
  Thacker, and O.M. Knio.
\newblock Global sensitivity analysis in an ocean general circulation model: a
  sparse spectral projection approach.
\newblock {\em Computational Geosciences}, 16(3):757--778, 2012.

\bibitem{LiIskandaraniLeHenaffEtAl16}
G.~Li, M.~Iskandarani, M.~Le~H{\'e}naff, J.~Winokur, O.P. Le~Ma{\^\i}tre, and
  O.M. Knio.
\newblock Quantifying initial and wind forcing uncertainties in the gulf of
  mexico.
\newblock {\em Computational Geosciences}, 20(5):1133--1153, 2016.

\bibitem{Namhata2016OladyshkinDilmoreEtAl16}
A.~Namhata, S.~Oladyshkin, R.M. Dilmore, L.~Zhang, and D.V. Nakles.
\newblock Probabilistic assessment of above zone pressure predictions at a
  geologic carbon storage site.
\newblock {\em Scientific reports}, 6:39536, 2016.

\bibitem{deman2016}
G.~Deman, K.~Konakli, B.~Sudret, J.~Kerrou, P.~Perrochet, and
  H.~Benabderrahmane.
\newblock Using sparse polynomial chaos expansions for the global sensitivity
  analysis of groundwater lifetime expectancy in a multi-layered
  hydrogeological model.
\newblock {\em Reliability Engineering \& System Safety}, 147:156--169, 2016.

\bibitem{SaadAlexanderianPrudhommeEtAl17}
B.~Saad, A.~Alexanderian, S.~Prudhomme, and O.M. Knio.
\newblock Probabilistic modeling and global sensitivity analysis for $ co\_2 $
  storage in geological formations: a spectral approach.
\newblock {\em Applied Mathematical Modelling}, 53:584--601, 2018.

\bibitem{DegasperiGilmore08}
A.~Degasperi and S.~Gilmore.
\newblock Sensitivity analysis of stochastic models of bistable biochemical
  reactions.
\newblock In {\em Formal Methods for Computational Systems Biology}, pages
  1--20. Springer, 2008.

\bibitem{navarro2016global}
M~Navarro~J., O.P. Le~Ma{\^\i}tre, and O.M. Knio.
\newblock Global sensitivity analysis in stochastic simulators of uncertain
  reaction networks.
\newblock {\em The Journal of Chemical Physics}, 145(24):244106, 2016.

\bibitem{Vohra:2014}
M.~Vohra, J.~Winokur, K.R. Overdeep, P.~Marcello, T.P. Weihs, and O.M. Knio.
\newblock Development of a reduced model of formation reactions in zr-al
  nanolaminates.
\newblock {\em Journal of Applied Physics}, 116(23):233501, 2014.

\bibitem{Sobol:2009}
I.M. Sobol' and S.~Kucherenko.
\newblock Derivative based global sensitivity measures and their link with
  global sensitivity indices.
\newblock {\em Mathematics and Computers in Simulation}, 79(10):3009--3017,
  2009.

\bibitem{Sobol:2010}
I.M. Sobol and S.~Kucherenko.
\newblock Derivative based global sensitivity measures.
\newblock {\em Procedia-Social and Behavioral Sciences}, 2(6):7745--7746, 2010.

\bibitem{Lamboni:2013}
M.~Lamboni, B.~Iooss, A.L. Popelin, and F.~Gamboa.
\newblock Derivative-based global sensitivity measures: general links with
  sobol' indices and numerical tests.
\newblock {\em Mathematics and Computers in Simulation}, 87:45--54, 2013.

\bibitem{Kucherenko:2009}
S.~Kucherenko, M.~Rodriguez-Fernandez, C.~Pantelides, and N.~Shah.
\newblock Monte carlo evaluation of derivative-based global sensitivity
  measures.
\newblock {\em Reliability Engineering \& System Safety}, 94(7):1135--1148,
  2009.

\bibitem{Kucherenko:2016}
S.~Kucherenko and B.~Iooss.
\newblock Derivative-based global sensitivity measures.
\newblock {\em Handbook of Uncertainty Quantification}, pages 1--24, 2016.

\bibitem{Kiparissides:2009}
A.~Kiparissides, S.S. Kucherenko, A.~Mantalaris, and E.N. Pistikopoulos.
\newblock Global sensitivity analysis challenges in biological systems
  modeling.
\newblock {\em Industrial \& Engineering Chemistry Research},
  48(15):7168--7180, 2009.

\bibitem{jameson1988aerodynamic}
A.~Jameson.
\newblock Aerodynamic design via control theory.
\newblock {\em Journal of scientific computing}, 3(3):233--260, 1988.

\bibitem{gunzburger2003perspectives}
M.D. Gunzburger.
\newblock {\em Perspectives in flow control and optimization}, volume~5.
\newblock Siam, 2003.

\bibitem{Borzi2011}
A.~Borz{\`\i} and V.~Schulz.
\newblock {\em Computational optimization of systems governed by partial
  differential equations}, volume~8.
\newblock SIAM, 2011.

\bibitem{AlexanderianPetraStadlerEtAl17}
A.~Alexanderian, N.~Petra, G.~Stadler, and O.~Ghattas.
\newblock Mean-variance risk-averse optimal control of systems governed by
  {PDEs} with random parameter fields using quadratic approximations.
\newblock {\em SIAM Journal on Uncertainty Quantification}, 5:1166--1192, 2017.

\bibitem{Efron:2004}
B.~Efron, T.~Hastie, I.~Johnstone, R.~Tibshirani, et~al.
\newblock Least angle regression.
\newblock {\em The Annals of statistics}, 32(2):407--499, 2004.

\bibitem{Tibshirani:1996}
R.~Tibshirani.
\newblock Regression shrinkage and selection via the lasso.
\newblock {\em Journal of the Royal Statistical Society. Series B
  (Methodological)}, pages 267--288, 1996.

\bibitem{Blatman:2008}
G.~Blatman and B.~Sudret.
\newblock Sparse polynomial chaos expansions and adaptive stochastic finite
  elements using a regression approach.
\newblock {\em Comptes Rendus M{\'e}canique}, 336(6):518--523, 2008.

\bibitem{Marelli:2014}
S.~Marelli and B.~Sudret.
\newblock Uqlab: A framework for uncertainty quantification in matlab.
\newblock In {\em Vulnerability, Uncertainty, and Risk: Quantification,
  Mitigation, and Management}, pages 2554--2563. 2014.

\bibitem{Blatman:2009}
G.~Blatman.
\newblock {\em Adaptive sparse polynomial chaos expansions for uncertainty
  propagation and sensitivity analysis}.
\newblock PhD thesis, Clermont-Ferrand 2, 2009.

\bibitem{Griewank:2008}
A.~Griewank and A.~Walther.
\newblock {\em Evaluating derivatives: principles and techniques of algorithmic
  differentiation}, volume 105.
\newblock Siam, 2008.

\bibitem{Morris:1993}
M.D. Morris, T.J. Mitchell, and D.~Ylvisaker.
\newblock Bayesian design and analysis of computer experiments: use of
  derivatives in surface prediction.
\newblock {\em Technometrics}, 35(3):243--255, 1993.

\bibitem{Yetter:1991}
R.A. Yetter, F.L. Dryer, and H.~Rabitz.
\newblock A comprehensive reaction mechanism for carbon
  monoxide/hydrogen/oxygen kinetics.
\newblock {\em Combustion Science and Technology}, 79(1-3):97--128, 1991.

\bibitem{Das:1996}
L.M. Das.
\newblock Hydrogen-oxygen reaction mechanism and its implication to hydrogen
  engine combustion.
\newblock {\em International Journal of Hydrogen Energy}, 21(8):703--715, 1996.

\bibitem{Safta:2011}
C.~Safta, H.N. Najm, and O.M. Knio.
\newblock Tchem-a software toolkit for the analysis of complex kinetic models.
\newblock {\em Sandia Report, SAND2011-3282}, 2011.

\end{thebibliography}

\end{document}